\documentclass[tighten,iop,apj]{emulateapj}
\usepackage[english]{babel}
\usepackage{amsmath}
\usepackage{amsfonts}
\usepackage{amssymb}
\usepackage{graphicx}
\usepackage{color}
\usepackage{natbib}
\usepackage[normalem]{ulem}
\usepackage[bookmarks=false, pdfnewwindow=true, colorlinks=true, linkcolor=cyan, citecolor=cyan, filecolor=cyan, urlcolor=cyan]{hyperref}

\shorttitle{Discrete Ordinates vs Monte Carlo Neutrino Transport}
\shortauthors{Richers, Nagakura, \emph{et al.}}

\begin{document}

\title{A detailed comparison of multi-dimensional Boltzmann neutrino transport methods in core-collapse supernovae}

\author{Sherwood Richers$^{1,*}$, Hiroki Nagakura$^1$, Christian D. Ott$^{2,1}$, Joshua Dolence$^3$, Kohsuke Sumiyoshi$^4$, \\ and Shoichi Yamada$^{5,6}$}

 \affiliation{$^1$TAPIR, Walter Burke Institute for Theoretical Physics, Mailcode 350-17, California Institute of Technology, Pasadena, CA 91125, USA}

\affiliation{$^2$ Center for Gravitational Physics and International Research Unit of Advanced Future Studies, Yukawa Institute for Theoretical Physics, Kyoto University, Kitashirakawa Oiwakecho, Sakyo-ku, Kyoto 606-8502, Japan}
 
 \affiliation{$^3$CCS-2, Los Alamos National Laboratory, P.O. Box 1663 Los Alamos, NM 87545}
 
 \affiliation{$^4$Numazu College of Technology, Ooka 3600, Numazu, Shizuoka 410-8501, Japan}

 \affiliation{$^5$Advanced Research Institute for Science \&
 Engineering, Waseda University, 3-4-1 Okubo,
 Shinjuku, Tokyo 169-8555, Japan}

 \affiliation{$^6$Department of Science and Engineering, Waseda
   University, 3-4-1 Okubo, Shinjuku, Tokyo 169-8555, Japan}
 \email{$^*$srichers@tapir.caltech.edu}

\begin{abstract}
  The mechanism driving core-collapse supernovae is sensitive to the
  interplay between matter and neutrino radiation. However, neutrino
  radiation transport is very difficult to simulate, and several
  radiation transport methods of varying levels of approximation are
  available. We carefully compare for the first time in multiple
  spatial dimensions the discrete ordinates (DO) code of Nagakura,
  Yamada, and Sumiyoshi and the Monte Carlo (MC) code {\tt Sedonu},
  under the assumptions of a static fluid background, flat spacetime,
  elastic scattering, and full special relativity. We find remarkably
  good agreement in all spectral, angular, and fluid interaction
  quantities, lending confidence to both methods. The DO method excels
  in determining the heating and cooling rates in the optically thick
  region. The MC method predicts sharper angular features due to the
  effectively infinite angular resolution, but struggles to drive down
  noise in quantities where subtractive cancellation is prevalent,
  such as the net gain in the protoneutron star and off-diagonal
  components of the Eddington tensor. We also find that errors in the
  angular moments of the distribution functions induced by neglecting
  velocity dependence are sub-dominant to those from limited
  momentum-space resolution. We briefly compare directly computed
  second angular moments to those predicted by popular algebraic
  two-moment closures, and find that the errors from the approximate
  closures are comparable to the difference between the DO and MC
  methods. Included in this work is an improved {\tt Sedonu} code,
  which now implements a fully special relativistic, time-independent
  version of the grid-agnostic Monte Carlo random walk approximation.
\end{abstract}

\keywords{supernovae: general---neutrinos---radiative transfer}

%%%%%%%%%%%%%%%%
% INTRODUCTION %
%%%%%%%%%%%%%%%%
\section{introduction}
\label{sec:introduction}

Most massive stars ($M\gtrsim 10M_\odot$) end their lives in a
cataclysmic core-collapse supernova (CCSN) explosion that releases
around $10^{51}\,\mathrm{erg}$ of kinetic energy and around
$10^{53}\,\mathrm{erg}$ of neutrino energy. The iron core begins to
collapse when it exceeds its effective Chandrasekhar mass as
degenerate electrons capture onto nuclei and photodissociation breaks
apart nuclei (e.g., \citealt{bethe:90}). Within a few tenths of a
second after the onset of collapse, the inner core becomes very
neutron-rich ($Y_e\sim0.3$) and exceeds nuclear densities
($\sim2.7\times10^{14}\,\mathrm{g\,cm}^{-3}$). At this point, the
strong nuclear force kicks in, dramatically stiffening the equation of
state (EOS) and abruptly stopping the collapse of the inner core
within a few milliseconds. The inner core then rebounds, sending a
shock wave through the supersonically infalling outer core. Neutrino
cooling removes energy from the matter under the shock and
photodissociation of heavy nuclei weakens the shock. The shock
subsequently stalls at around $150\,\mathrm{km}$ as it is lacks
pressure support from below to overcome the ram pressure of the
accreting outer stellar core.

Understanding the mechanism that revives the shock's outward progress
and results in a CCSN is presently the main target of CCSN theory. The
canonical theory is the neutrino mechanism \citep{bethe:85}, whereby
neutrinos emitted from the dense inner core pass through the matter
below the shock, depositing enough thermal energy to revive the shock
via thermal support and by driving turbulence (e.g.,
\citealt{janka:01,burrows:13a,mueller:16b}). However, the strongly
nonlinear dynamics in this stage is inexorably coupled to a variety of
microphysical processes. In particular, it is very sensitive to
properties of the neutrino field passing through the matter. During
the stalled shock phase, the star delicately straddles the line
between explosion and total collapse, so small differences in how the
neutrinos interact with the matter can be the difference between an
explosion and a dud (e.g., \citealt{janka:01, murphy:08, oconnor:11,
  melson:15b, couch:15a, burrows:16}).

Computation has become the primary tool for studying these nonlinear
processes, as it allows us to see detailed dynamics and make
observable predictions (electromagnetic radiation, neutrinos,
gravitational waves) under the assumptions imposed by the
model. However, computational techniques and resources are still too
primitive to allow for simulations complete with all of the required
fidelity and involved physics. In general, simulations of CCSNe
require a three-dimensional general-relativistic (GR) treatment of
magnetohydrodynamics, neutrino radiation transport, and a
microphysical equation of state (EOS) (e.g.,
\citealt{kotake:12snreview,janka:12a,ott:16}). The simulations also require
sufficient resolution or sub-grid modeling to capture everything from
global dynamics to 100 meter-scale or smaller turbulence
\citep{radice:16a,ott:16}.

Deep in the inner core, neutrinos are trapped and form an isotropic
thermal distribution that slowly diffuses out. Outside the shock, the
neutrinos are free-streaming and move only radially outward. Though
radiation transport methods are constructed to simulate these limits
well, the intermediate semi-transparent region is challenging to
accurately simulate. This region is responsible for most of the
dynamics that support the shock's progress due to neutrino heating. In
addition, the neutrino opacity scales approximately as the square of
the neutrino energy, causing the energy deposition rate and the
location of the transition from trapped to free-streaming to depend
sensitively on neutrino energy. Hence, we require a means of
simulating neutrinos of many energies in all regions of a CCSN.

A full treatment of classical neutrino radiation requires evolving the
neutrino distribution function of each neutrino species according to
the seven-dimensional Boltzmann equation {\textcolor{black}{(e.g., \citealt{lindquist:66,ehlers:93,mihalas:99})}}
(three spatial dimensions, three momentum dimensions, time), which
presents a significant computational challenge. A wide variety of
methods have been used to capture the most important aspects of
neutrino transport through the supernova that can be broadly
categorized as either phenomenological, deterministic, or
probabilistic methods. Though some methods are definitively more
accurate than others, there is always a trade-off between efficiency
and accuracy.

Phenomenological approaches include the light bulb scheme and neutrino
leakage and only very approximately account for neutrino
effects. These schemes are very efficient, making them very conducive
to parameter studies. In the light bulb scheme (e.g.,
\citealt{bethe:85,janka:96,ohnishi:06,murphy:08}) the luminosity and
temperature of each neutrino species are simply input parameters. All
heating rates are based on this parameter and cooling rates are
estimated based on an approximate optical depth. The inner light bulb
boundaries have also been combined with gray transport schemes in the
semi-transparent and transparent regimes \citep{scheck:06}. In the
leakage scheme, an approximate neutrino optical depth at each point is
calculated, and this is used to set the cooling rate at each point
\citep{ruffert:96,rosswog:03b}. Neutrino heating can be included
approximately by assuming that the neutrino luminosity through a given
point is determined by the energy leaking radially outward from below
\citep{oconnor:10,fernandez:13,perego:16}.

Approximate deterministic methods solve a simplified version of the
Boltzmann equation in order to make a more tractable problem. The
isotropic diffusion source approximation (IDSA) method evolves an
isotropic trapped component and a free-streaming component of the
distribution function \citep{liebendoerfer:09}. In truncated moment
methods, the distribution function is discretized into an infinite
list of angular moments, only the first few of which are directly
evolved. Flux-limited diffusion (FLD,
\citealt{levermore:81,mihalas:82,pomraning:83,castor:04,krumholz:07})
is a one moment method that evolves only the zeroth moment of the
distribution function (energy density) and requires a closure relation
to estimate the first moment (flux). However, FLD fails to capture
much of the angular information about the distribution function and
tends to smooth out angular variations (e.g.,
\citealt{janka:92c,burrows:00,liebendoerfer:04,ott:08,zhang:13}).  In
the two-moment method \citep{pomraning:69, anderson:72, thorne:81,
  dubroca:99, audit:02, shibata:11, vaytet:11, cardall:13rad}, the
zeroth and first moments are evolved, and a closure relation is
required to estimate the second moment (pressure tensor) and complete
the system of equations. The closure can be provided by some ad-hoc
analytical function (e.g., \citealt{smit:00,murchikova:17} and
references therein). This is also known as the M1 method, though M1
confusingly refers to a specific closure as well
\citep{levermore:84,dubroca:99}. The closure can also be more
accurately determined using a direct solution of the Boltzmann
equation, referred to as a variable Eddington tensor (VET) method
(e.g., \citealt{stone:92,hayes:03}).

In discrete ordinates (DO or $S_n$) methods, the distribution function
is discretized into angular bins, each of which is directly evolved
(e.g., \citealt{pomraning:69,mihalas:99} and references therein). In
spherical harmonic ($P_n$) methods, the distribution function is
decomposed in terms of spherical harmonics, and a small number of
these are evolved (e.g.,
\citealt{pomraning:73,radice:13rad}). Finally, fully spectral methods
in all six space-momentum dimensions have been applied to stationary
neutrino transport calculations \citep{peres:14}.

Monte Carlo (MC) radiation transport \citep{fleck:71, fleck:84,
  densmore:07, abdikamalov:12} is a probabilistic method that samples
the trajectories of a finite number of individual neutrinos and
assumes their behavior is representative of the rest of the bulk
neutrino behavior. \cite{tubbs:78} applied MC methods to neutrino
transport for the first time to study neutrino-matter equilibration in
an infinite uniform medium. MC transport has been long used in 1D
steady-state transport calculations \citep{janka:89c, janka:91,
  janka:92c, yamada:99, keil:03, abdikamalov:12}, though the code of
\cite{abdikamalov:12} was also designed to perform time-dependent
calculations. \cite{richers:15} performed steady-state MC transport
calculations on 2D snapshots of accretion disks from neutron star
mergers, though the optical depths were much lower than in the CCSN
context.

There is much more to a radiation transport code than the broad
classes of methods mentioned above. Detailed Boltzmann transport using
either DO or VET methods in dynamical simulations has been achieved in
one \citep{mezzacappa:93b, yamada:99, burrows:00, rampp:02,
  roberts:12b} and two \citep{livne:04,ott:08,nagakura:17a} spatial
dimensions, but three dimensional calculations are presently only
possible when the fluid is assumed to be stationary
\citep{sumiyoshi:12, sumiyoshi:15}. The local two-moment method is
used in 1D \citep{oconnor:15a}, 2D \citep{oconnor:15b,just:15}, and 3D
(\citealt{sekiguchi:15,roberts:16c,foucart:16,kuroda:16}, see also
\citealt{mueller:15}) core collapse and neutron star merger
simulations. FLD is also popular in 2D simulations
\citep{dessart:06pns, swesty:09, zhang:13}. Various versions of the
ray-by-ray (RbR) approximation can also be used to extend a 1D
transport method to two \citep{bhf:95, buras:06a, mueller:10, suwa:10,
  pan:16} or three \citep{hanke:13, takiwaki:12, lentz:15, melson:15a}
spatial dimensions in an efficient manner by making transport along
individual radial rays nearly independent of other rays and/or by
solving a single spherically averaged 1D transport problem. Fully
general relativistic neutrino radiation hydrodynamics simulations are
now possible \citep{janka:91, yamada:99, liebendoerfer:01,
  sumiyoshi:05, mueller:10, shibata:12, kuroda:12, oconnor:15a,
  foucart:16, roberts:16c, kuroda:16} and many codes incorporate
general relativistic effects with various levels of approximation
(e.g., \citealt{mueller:10, oconnor:15b,
  skinner:16}). \textcolor{black}{{Special-relativistic effects can be accounted for in
full generality \citep{mueller:10, shibata:12, richers:15, foucart:16,
  nagakura:17a} or by using only up to $\mathcal{O}(v/c)$ terms (e.g.,
\citealt{rampp:02, lentz:15, just:15, dolence:15,
  skinner:16}).}} Even if the transport method is equivalent,
simulations differ in how neutrino-matter and neutrino-neutrino
interactions are treated (e.g., \citealt{lentz:12b}). In short, each
piece of relevant physics can be simulated accurately, but 3D
simulations containing all pieces remain a goal that has not yet been
achieved.

Any computational method is an approximation of reality, and every
method has strengths and weaknesses. It is therefore expected that
computations performed by different codes should arrive at different
solutions, though they should converge to the physical answer with
increasing simulation fidelity. Understanding the weaknesses of a
given method is a prerequisite to interpreting the physical meaning of
simulation results. It is standard practice to test that codes produce
known solutions to simple problems and to perform self-convergence
tests to ensure that results are not mistakes or numerical
artifacts. However, even with these practices in place, different
codes produce different results and independent verification is
required to help determine which features of each are realistic
\citep{calder:02}.

Several works in the past have evaluated the accuracy of a low-order
method like FLD or two moment transport by comparing with a high-order
DO, MC, or VET method (e.g., \citealt{janka:92c, mezzacappa:93b,
  messer:98, burrows:00, liebendoerfer:04, ott:08}). However, in these
comparisons, the low-order method is not expected to converge to the
same result as the high-order method, which does not help to verify
the high-order method. There have been few detailed comparisons
between high-order methods that solve the same full Boltzmann equation
(i.e., VET, DO, and MC methods), and none in more than one spatial
dimension. \cite{yamada:99} compared the results of a new DO
implementation to the MC code of \cite{janka:91} in 1D GR snapshots of
CCSN simulations, but ignored fluid motion. \cite{liebendoerfer:05}
performed a comparison of dynamical 1D CCSN simulations using the GR
DO code Agile-BOLTZTRAN \citep{liebendoerfer:04} with the Newtonian
VET code VERTEX-PROMETHEUS \citep{rampp:02}. They found very good
agreement once an effective potential was introduced to
VERTEX-PROMETHEUS to account for GR effects \citep{marek:06}. Several
groups have since used the results of \cite{liebendoerfer:05} as a
standard for comparison (e.g., \citealt{sumiyoshi:05,mueller:10,
  suwa:11a,lentz:12a, lentz:12b, oconnor:13, oconnor:15a, just:15b,
  suwa:16}). With the recent arrival of the many advanced
multi-dimensional neutrino radiation hydrodynamics codes mentioned
previously, continued independent verification is essential to
interpreting simulation results.

In this paper, we perform the first detailed multi-dimensional
comparison between fully special relativistic Boltzmann neutrino
transport codes using a DO neutrino radiation hydrodynamics code
\citep{nagakura:17a} (hereafter NSY) and the MC radiation transport
code \texttt{Sedonu} \citep{richers:15}. We make the time-independent
comparisons on spherically symmetric (1D) and in axisymmetric (2D)
snapshots from CCSN simulations at around $100\,\mathrm{ms}$ after
core bounce. Both codes are carefully configured to calculate the full
steady-state neutrino distribution function from first principles in
as similar a manner as possible. We find remarkably good agreement in
all spectral, angular, and fluid interaction quantities, lending
confidence to both methods. The MC method predicts sharper angular
features due to the effectively infinite angular resolution, but
struggles to drive down noise in quantities where subtractive
cancellation is prevalent (e.g., net gain within the protoneutron star
and off-diagonal components of the Eddington tensor). We test the
importance of accounting for fluid velocities by setting all
velocities to zero and find that the differences induced are much
smaller than the errors due to finite momentum-space resolution. We
compare directly computed second angular moments to those predicted by
popular two-moment closures, and find that the error from the
approximate closure is comparable to the difference between the DO and
MC methods.
  
This paper is organized as follows. In Section~\ref{sec:transmethods},
we review the discrete ordinates, Monte Carlo, and two moment
methods. We present the results of the transport method comparisons in
spherical symmetry in Section~\ref{sec:1Dcomparison} and in axial
symmetry in Section~\ref{sec:2Dcomparison}.  We summarize our
conclusions in Section~\ref{sec:conclusions}.  The MC transport code
{\tt Sedonu} is publicly available at
\url{https://bitbucket.org/srichers/sedonu} and the results obtained
in this study from both transport codes are available at
\url{https://stellarcollapse.org/MCvsDO}.

%%%%%%%%%%%
% METHODS %
%%%%%%%%%%%
\section{Numerical Methods of Neutrino Transport}
\label{sec:transmethods}
The transport of classical neutrinos is described in general by the
Boltzmann equation \textcolor{black}{{(e.g., \citealt{lindquist:66,ehlers:93,mihalas:99})}}:
\begin{equation}
  \frac{df}{d\lambda} = \left[\frac{\partial f}{d\lambda}\right]_\mathrm{em-abs} + \left[\frac{\partial f}{d\lambda}\right]_\mathrm{scat} + \left[\frac{\partial f}{d\lambda}\right]_\mathrm{pair}\,\,,
  \label{eq:boltzmann}
\end{equation}
where $\lambda$ is an affine parameter. In a coordinate basis, the
geodesic equation gives
\begin{equation}
  \frac{d f}{d \lambda} = p^\alpha \frac{\partial f}{\partial x^\alpha} - \Gamma^\alpha_{\mu\nu}p^\mu p^\nu \frac{\partial f}{\partial p^\alpha} \,\,,
\end{equation}
where $p^\alpha$ are the neutrino four-momenta, $x^\alpha$ are the four
spacetime coordinates, and $\Gamma^\alpha_{\mu\nu}$ are Christoffel
symbols of the spacetime metric. In an orthogonal coordinate system in
flat spacetime, the distribution function $f$ of each neutrino species
is defined as
\begin{equation}
  f(\mathbf{x}, \mathbf{\Omega}, \epsilon, t) = \frac{dN}{dV\,d\mathbf{\Omega}\,d\epsilon}\,\,,
  \label{eq:dfunc}
\end{equation}
where $N$ is the number of neutrinos, $dV(\mathbf{x})$ is the volume
element at position $\mathbf{x}$, $t$ is the time, $\mathbf{\Omega}$
is the neutrino direction, and $\epsilon$ is the neutrino energy. The
three source terms on the right hand side of
Equation~\ref{eq:boltzmann} are interaction terms from emission and
absorption, scattering, and pair processes, respectively. The neutrino
propagation is generally calculated in reference to coordinates
defined in the lab frame, but interactions between matter and
neutrinos are formulated in the fluid rest frame (a.k.a.\ the comoving
frame). It is thus important to very carefully keep track of the frame
in which various quantities are defined. This is consistent with
widely used conventions in the relativistic neutrino transport
community (e.g., \citealt{shibata:11,cardall:13rad}).

The interaction terms are all local and formulated in a frame comoving
with the underlying fluid. The emission and absorption terms are the
simplest, as they depend linearly on the distribution function of a
given neutrino species. The scattering term in general depends
quadratically on the distribution function of a given species, since
neutrinos are fermions and the reaction is inhibited by final-state
neutrino blocking. However, under the assumption of isoenergetic
scattering, it reduces to a linear dependence (see
Appendix~\ref{app:sourceterms}). The pair term, which includes
neutrino pair annihilation and creation and neutrino bremsstrahlung,
depends on the product of the distribution functions of the species
and anti-species. Our static transport calculations solve for the $f$
that satisfies $\partial f/\partial t=0$.

There are six species of neutrinos in the standard model corresponding
to the six leptonic species ($\nu_e$,$\nu_\mu$,$\nu_\tau$ and their
anti-particles). Electron neutrinos and anti-neutrinos interact with
nucleons via both charged current and neutral current processes, while
heavy lepton neutrinos interact only via neutral current
processes. This makes each heavy lepton neutrino species individually
less impactful than electron anti/neutrinos and making all four
species behave very similarly. In light of this, we simulate $\nu_e$
and $\bar{\nu}_e$ individually, but group all of the heavy-lepton
neutrinos into a single simulated species $\nu_x$ for computational
efficiency.

Neutrino interaction rates depend on the properties of the fluid
through which they traverse. In this study, we use the non-hyperonic
equation of state (EOS) of \cite{hshen:11} to determine the abundances
and chemical potentials of each constituent (i.e., leptons, nucleons,
and nuclei) given the fluid density, temperature, and electron
fraction.  We consider the following minimum but essential sets of
neutrino-matter interactions in the postbounce phase of CCSNe:
\begin{equation*}
  \begin{aligned}
e^{-} + p &\longleftrightarrow& \nu_{e} + n\,\,,\\
e^{+} + n &\longleftrightarrow& \bar{\nu_{e}} + p\,\,,\\
\nu (\bar{\nu})  + N &\longleftrightarrow& \nu (\bar{\nu}) + N\,\,,\\
e^{-} + e^{+} &\longleftrightarrow&  \nu + \bar{\nu}\,\,,\\
N + N &\longleftrightarrow& N + N + \nu + \bar{\nu}\,\,.\,
\end{aligned}
\end{equation*}
where $N \in \{n,p\}$. From top to bottom, these processes are
electron capture by free protons, positron capture by free neutrons,
isoenergetic scattering with nucleons, electron-positron pair
annihilation, and nucleon-nucleon bremsstrahlung, along with each of
their inverse reactions.

Though a multitude of phenomenological, approximate, and exact
transport methods exist in the literature, we will focus on three of
them. The discrete ordinates method (DO, Section~\ref{sec:DO}) and the
Monte Carlo (MC, Section~\ref{sec:MC}) methods described here both
solve the Boltzmann equation directly in all three momentum dimensions
and multiple spatial dimensions, and so should converge to the same
physical result. We also investigate how well approximate closure
relations in the two-moment method (Section~\ref{sec:twomoment})
compare to the solutions computed by DO and MC calculations.

\subsection{Discrete Ordinates}
\label{sec:DO}

The DO Boltzmann code of Nagakura, Sumiyoshi, and Yamada (hereafter
NSY) is a grid-based multi-dimensional neutrino radiation
hydrodynamics code that solves the conservative form of
Equation~\ref{eq:boltzmann} in the language of the 3+1 formulation of
general relativity (GR). The numerical method is essentially the same
as described by \cite{sumiyoshi:12}, though it has since been extended
to account for special relativistic effects as has been coupled with
Newtonian hydrodynamics \citep{nagakura:14,nagakura:17a}. The newest
version of this code was recently applied to axisymmetric CCSN
simulations in \cite{nagakura:17b}.

The neutrino distribution function $f$ is discretized onto a
spherical-polar spatial grid described by radius $r$, polar angle
$\theta$, and azimuthal angle $\phi$.  The radial grid is constructed
so as to provide good resolution where the density gradient is
large. The radial mesh spacing is set to $\Delta r = 300\,\mathrm{m}$
at the center and decreases with increasing radius up to the location
of the steepest density gradient at $r = 10\,\mathrm{km}$, where
$\Delta r = 104\,\mathrm{m}$. For $ r \geq 10\,\mathrm{km}$, the
spacing increases by $1.7 \%$ per zone up to $r=500\,\mathrm{km}$. For
$r \geq 500\,\mathrm{km}$, the spacing increases by $3.8 \%$ per zone
up to the outer boundary of $r=5000\,\mathrm{km}$. This results in 384
radial grid zones over the entire domain. The spatial angular grid is
set to 128 Gaussian quadrature points from $0 \leq \theta \leq
\pi$. At each spatial location, $f$ is discretized onto a
spherical-polar momentum space grid described by neutrino energy
$\epsilon$, neutrino polar angle $\bar{\theta}$ (where
$\bar{\theta}=0$ is in the radial direction), and neutrino azimuthal
angle $\bar{\phi}$. The first bin of the neutrino energy grid extends
over $0-2\,\mathrm{MeV}$ in the 1D\_1x and 2D calculations,
$0-1\,\mathrm{MeV}$ in the 1D\_2x calculations, and
$0-0.5\,\mathrm{MeV}$ in the 1D\_4x calculations. The rest of the
energy bins are logarithmically spaced from $2\,\mathrm{MeV}$ to
$300\,\mathrm{MeV}$.  The number of energy and direction bins used in
each simulation is listed in Table~\ref{tab:SimulationList}.

The NSY code treats the advection terms in the GR Boltzmann equation
semi-implicitly. Both advection and collision terms are implemented
self-consistently by using a mixed-frame approach with separate
momentum-space grids in the lab and comoving frames. See
Appendix~\ref{app:boltzmann} and \cite{nagakura:17a} for
implementation details.

Though the NSY code is capable of evolving coupled neutrino radiation
hydrodynamics, we restrict the capability of the code in this study to
evolve only the radiation field on top of a fixed fluid background
with a flat spacetime metric until a nearly steady-state solution is
reached.  As we list in Table~\ref{tab:SimulationList}, the maximum
time variability in the energy density at any spatial location
relative to the value averaged over $1\,\mathrm{ms}$ is less than
$0.1\%$, which is significantly smaller than difference between DO and
MC results.

\subsection{Monte Carlo}
\label{sec:MC}

The MC method for radiation transport is a probabilistic
implementation of a reformulated Equation~\ref{eq:boltzmann}, making
it fundamentally different from the DO method. We employ the {\tt
  Sedonu} MC neutrino transport code \citep{richers:15} to solve for
equilibrium neutrino radiation fields and neutrino-matter interaction
rates. The neutrino radiation field of each neutrino species is
discretized into neutrino packets, each with some total packet energy
$E_p$ representing a large number of neutrinos at the same location
$\mathbf{x}$ with the same individual energy $\epsilon$ and the same
direction $\mathbf{\Omega}$. The neutrino motion itself is always
computed in the lab frame, and special relativistic effects are
accounted for by explicitly Lorentz transforming into and out of the
comoving frame for interactions with the background fluid.

Neutrinos are emitted from the fluid within each grid cell at a random
location, with an isotropically random direction in the comoving
frame, and with a random comoving energy $\epsilon$ according to the
energy-dependent emissivity. The comoving packet energies are set such
that the grid cell emits with a luminosity determined by the
energy-integrated emissivity.

Each particle moves linearly in a series of discrete steps in the lab
frame. The size of each step is the minimum of a random distance
determined by the scattering opacity $\kappa_s$ and the grid cell
distance $l_\mathrm{grid}$. $l_\mathrm{grid}$ is set to the maximum of
the distance to the grid cell boundary and 1\% of the grid cell's
smallest dimension to prevent neutrinos from getting stuck at cell
boundaries. After each step, the packet energy is diminished by a
factor of $\exp(-\kappa_a l)$, where $\kappa_a$ is the absorption
opacity and $l$ is the length of the step. When the packet undergoes
an elastic scatter, a new direction is chosen isotropically randomly
in the comoving frame. The absorption and scattering opacities are
re-evaluated when the neutrino enters a new grid cell. If the cell is
optically thick to neutrinos, we use a time-independent relativistic
version of the Monte Carlo random walk approximation \citep{fleck:84}
to allow the neutrino to make a large step through many effective
isotropic elastic scatters (see Appendix~\ref{app:montecarlo}).

The neutrino heating rates and radiation field quantities output by
\texttt{Sedonu} are the steady-state quantities by construction, and require no
concept of relaxation time. Upon emission, each neutrino packet energy
$E_p$ is set assuming an arbitrary emission time of $\delta
t_\mathrm{emit}$ and is allowed to propagate until it leaves the
simulation domain. After each step, \texttt{Sedonu} accumulates $\bar{E}_p l$
in the radiation field energy-direction bins, where $l$ is again the
distance the packet moves and $\bar{E}_p$ is the packet energy
averaged over the step. \texttt{Sedonu} additionally accumulates the amount of
the packet energy that is absorbed into the fluid. The steady-state
radiation energy density is obtained by normalizing the accumulated
radiation field by $V c\delta t_\mathrm{emit}$, where $V$ is the cell
volume. The steady-state comoving frame neutrino specific heating rate
is obtained by normalizing the deposited energy by $\rho V \delta
t_\mathrm{emit}$, where $\rho$ is the fluid rest density. This method
also provides a large speedup over time-dependent Monte Carlo
radiation transport for time-independent problems.

In this study, we perform the transport on 1D and 2D fluid grids in
spherical-polar coordinates that are identical to those employed by
the NSY code. We tally the radiation field in two different ways. In
the first, we use energy and direction bins identical to those used by
the NSY code. The data output is thus discretized, even though the
neutrinos themselves are always transported through continuous space
and are not influenced by any grid structure except in evaluating the
opacities. In the second way (``native''), we accumulate neutrino
energy directly into angular moments without any reference to a
discrete direction grid, though we still use the same energy bins. The
version of \texttt{Sedonu} used in this paper is fundamentally the same as that
used in \cite{richers:15}, except that it includes the Monte Carlo
random walk approximation for regions of large optical depth, the
native moment prescription, and various performance and usability
upgrades. \texttt{Sedonu} is open source and available at
\url{https://bitbucket.org/srichers/sedonu}.

\section{{{Eddington Tensor}} Analysis}
\label{sec:twomoment}

The so-called local two-moment transport method \citep{pomraning:69,
  anderson:72, thorne:81, shibata:11, cardall:13rad} is the current
state of the art method for time-dependent multi-dimensional
simulations of neutrino radiation hydrodynamics (e.g.,
\citealt{oconnor:15b, just:15b, foucart:16a, roberts:16c, kuroda:16,
  radice:17b})\footnote{Higher-order transport calculations (e.g.,
  \citealt{mueller:10}) are used in multiple dimensions via the
  ray-by-ray approximation (e.g., \citealt{bhf:95,buras:06a}), but we
  do not address these in this paper.}. In the two-moment method,
Equation~\ref{eq:boltzmann} is again reformulated, this time in terms
of specific moments of the distribution function. In an orthonormal
coordinate system, these moments are defined by
\begin{equation}
  \begin{aligned}
    \label{eq:moments}
    E_{\epsilon}(\mathbf{x},\epsilon)      &= \int d\mathbf{\Omega} \epsilon f(\mathbf{\Omega},\epsilon)\,\,,\\
    F^i_{\epsilon}(\mathbf{x},\epsilon)    &= \int d\mathbf{\Omega} \epsilon f(\mathbf{\Omega},\epsilon)[\mathbf{\Omega}\cdot\mathbf{e}_{(i)}]\,\,, \\
    P^{ij}_{\epsilon}(\mathbf{x},\epsilon) &= \int d\mathbf{\Omega} \epsilon f(\mathbf{\Omega},\epsilon) [\mathbf{\Omega}\cdot\mathbf{e}_{(i)}][\mathbf{\Omega}\cdot\mathbf{e}_{(j)}]\,\,,\\
    ...
  \end{aligned}
\end{equation}
where $\epsilon$ is the neutrino energy and $\mathbf{e}_{(i)}$ are the
basis vectors. The ellipsis denotes that there is an infinite list of
moments that can be used to reconstruct the two-dimensional angular
dependence of the distribution function, much like an infinite list of
terms in a Taylor series can be used to reconstruct a one-dimensional
function. Each of these specific moments
($M_\epsilon \in \{E_\epsilon,F_\epsilon^i, P_\epsilon^{ij},...\}$) can be
integrated in energy according to
\begin{equation}
  M = \frac{1}{(hc)^3}\int d\left(\frac{\epsilon^3}{3}\right) M_{\epsilon}
\end{equation}
to get the total neutrino energy density, energy flux, etc..

In the two-moment method, only the first two moments are evolved and
are assumed to provide a good enough representation of the full
distribution function. The evolution equations for each moment depend
on higher-order moments. In a \emph{local} two-moment scheme,
the pressure tensor and higher-order moments are estimated based on
the energy density and flux at the same spatial location. This
estimate is referred to as a closure relation, many of which have been
proposed based on various motivations (e.g.,
\citealt{smit:00,murchikova:17} and references therein).

In this study, we do not perform two-moment radiation transport, but
rather compare the radiation pressure tensor predicted by approximate
closures to the actual radiation pressure tensor output from the MC
and DO calculations. We consider three popular approximate closure
relations based on different physical motivations: the maximum entropy
closure of \cite{minerbo:78} in the classical limit (Minerbo), the
isotropic rest-frame closure of \cite{levermore:84} (Levermore), and
the closure of \cite{janka:91} in the form presented in
\cite{just:15b} that is empirically based on MC calculations of
neutrino transport in protoneutron stars. In all three cases, the
pressure tensor is expressed as an interpolation between optically
thick and thin limits where the solution is known analytically:
\begin{equation}
    P^{ij}_\epsilon = \frac{3\chi_\epsilon-1}{2}P^{ij}_{\epsilon,\mathrm{thin}} + \frac{3(1-\chi_\epsilon)}{2}P^{ij}_{\epsilon,\mathrm{thick}}\,\,,
\end{equation}
where
\begin{equation}
  \begin{aligned}
    P^{ij}_{\epsilon,\mathrm{thin}} &= E_\epsilon \frac{F_\epsilon^i F_\epsilon^j}{\mathbf{F}_\epsilon\cdot \mathbf{F}_\epsilon}\,\,,\\
    P^{ij}_{\epsilon,\mathrm{thick}} &= \frac{E_\epsilon}{3} I^{ij} + \mathcal{O}\left(\frac{v}{c}\right)\,\,,
  \end{aligned}
\end{equation}
and $I$ is the $3\times3$ identity matrix. In our analysis, we ignore
the $\mathcal{O}(v/c)$ term for simplicity. The different closure
relations are defined by
\begin{equation}
  \label{eq:closures}
  \begin{aligned}
    \chi_{\epsilon,\mathrm{Minerbo}} &= \frac{1}{3} + \frac{2}{15}\zeta_\epsilon^2\left(3 - \zeta_\epsilon + 3\zeta_\epsilon^2\right)\,\,,\\
    \chi_{\epsilon,\mathrm{Levermore}} &= \frac{3+4\zeta_\epsilon^2}{5+2\sqrt{4-3\zeta_\epsilon^2}}\,\,,\\
    \chi_{\epsilon,\mathrm{Janka}} &= \frac{1}{3}\left(1+\zeta_\epsilon^{1.31} + 1.5\zeta_\epsilon^{3.56}\right)\,\,.
  \end{aligned}
\end{equation}
$\zeta_\epsilon=\sqrt{\mathbf{F}_\epsilon\cdot \mathbf{F}_\epsilon /
  E_\epsilon^2}$ is referred to as the the flux factor. $P^{ij}/E$ is
referred to as the Eddington tensor, and in the context of
spherically-symmetric radiation transport, $P^{rr}/E$ is referred to
as the Eddington factor.

%%%%%%%%%%%%%%
% 1D RESULTS %
%%%%%%%%%%%%%%
\section{Transport Comparison in Spherical Symmetry}
\label{sec:1Dcomparison}

\begin{deluxetable*}{lccccc}
  \tablecolumns{6}
  \tablecaption{List of Calculations}
    \tablehead{    \colhead{Problem Name} & \colhead{Special Relativity} & \colhead{Spatial Resolution} & \colhead{Angular Resolution} & \colhead{MC Particles} & \colhead{DO $\delta E_{\rm max}$} \\
         &            & \colhead{$r\times\theta\times\epsilon$} & \colhead{$\bar{\theta}\times\bar{\phi}$} & \colhead{($\times 10^9$)} & }
    \startdata
    \sidehead{1D Calculations}
    1D\_1x         & yes & $384\times1\times20$ & $10\times1\phantom{0}$ & 1.82 & $2.40 \times 10^{-4\phantom{0}}$ \\
    1D\_2x         & yes & $384\times1\times40$ & $20\times1\phantom{0}$ & 2.33 & $5.50 \times 10^{-8}\phantom{0}$ \\
    1D\_4x         & yes & $384\times1\times80$ & $40\times1\phantom{0}$ & 2.67 & $7.40 \times 10^{-8\phantom{0}}$ \\
    1D\_4x\_nonrel & no  & $384\times1\times80$ & $40\times1\phantom{0}$ & -- & $9.80 \times 10^{-8\phantom{0}}$ \\
    1D\_4x\_native & yes & $384\times1\times80$ & --                     & 2.96 & --\\
    1D\_4x\_native\_nonrel & no  & $384\times1\times80$ & -- & 3.25 & --\\
    \sidehead{2D Calculations}
    2D\_LR         & yes & $270\times128\times20$ & $10\times6\phantom{0}$ & --  & $ 7.00 \times 10^{-4} $ \\
    2D\_LR\_nonrel & no  & $270\times128\times20$ & $10\times6\phantom{0}$ & --  & $ 5.84 \times 10^{-4}$ \\
    2D\_HR         & yes & $270\times128\times28$ & $14\times10$           & -- &  $ 4.84 \times 10^{-4} $ \\
    2D\_HR\_native & yes & $270\times128\times28$ & --                     & 63.4 & -- \\
    \enddata

    \tablecomments{We list the numerical details of
      each calculation presented in this paper. The leftmost column
      gives the name of each steady-state problem that is solved by
      one or both of {\tt Sedonu} and the NSY code. The Special
      Relativity column indicates whether special relativistic effects
      are taken into account. The Resolution column describes the
      spatial resolution of the fluid data and the number of neutrino
      energy bins. The Angular Resolution column describes the number
      of discrete angular bins in the neutrino momentum space angular
      discretization of the distribution function. In the ``native''
      MC calculations, neutrinos are accumulated directly into angular
      moments ($E_\epsilon$, $F^r_\epsilon$, $F^\theta_\epsilon$,
      $P^{rr}_\epsilon$, $P^{r\theta}_\epsilon$,
      $P^{\theta\theta}_\epsilon$, $P^{\phi\phi}_\epsilon$) without
      making reference to any discrete angular representation of the
      distribution function. The final two columns are indicative of
      the fidelity of the simulation. The MC Particles column shows
      the number of MC neutrino packets that we simulate. The DO
      $\delta E_{\rm max}$ measures the time variability in the steady
      state solution in DO calculations.  $\delta E_{\rm max}$ is
      defined as the maximum difference from time-average energy
      density during the final $1\,\mathrm{ms}$ of the simulation in
      $r<500$km.}
  \label{tab:SimulationList}
  \end{deluxetable*}
  
We perform several spherically symmetric (1D) steady-state neutrino
transport calculations using different momentum-space treatments
listed in Table~\ref{tab:SimulationList}. In the 1D\_1x, 1D\_2x, and
1D\_4x simulations, the neutrino radiation field is discretized
into the number of angular and energy bins described in the ``Spatial
Resolution'' and ``Angular Resolution'' columns. They differ only by
the number of spatial and energy bins, and are all run in each the
NSY code and {\tt Sedonu}. The 1D\_4x\_nonrel calculation (performed
only by the NSY code) is identical to the 1D\_4x calculation, except
with all velocities set to zero. In the 1D\_4x\_native calculation
performed only by {\tt Sedonu}, MC particles are accumulated directly
into angular moments rather into angular bins. The
1D\_4x\_native\_nonrel calculation is identical to the 1D\_native
calculation, but with all velocities set to zero. Note that length
contraction causes the simulations that include relativistic effects
to have a slightly larger total rest mass, but only by
$2\times10^{-5}\,M_\odot$. Throughout this section, we primarily
compare the highest-resolution DO calculation (1D\_4x) to the
highest-resolution native-moment MC calculation (1D\_4x\_native). We
use the lower resolution calculations (1D\_1x and 1D\_2x) in
resolution comparisons.

\begin{figure}
\includegraphics[width=\linewidth]{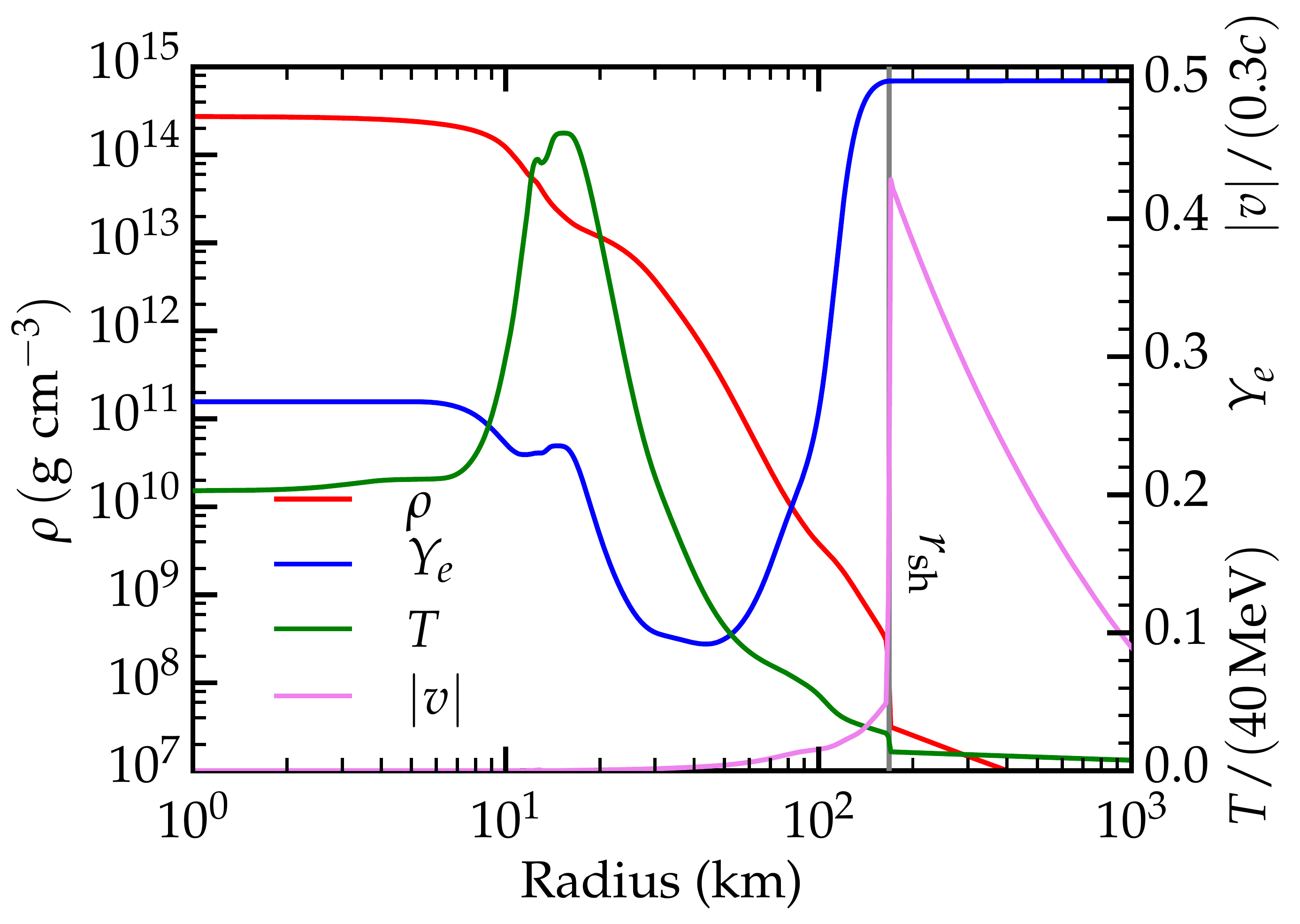}
\caption{\textbf{1D Fluid Properties.} This fluid snapshot from a 1D
  dynamical simulation in the NSY code at $100\,\mathrm{ms}$ after
  bounce is the background on which we solve the 1D steady-state
  radiation transport problem. The background fluid density $\rho$
  (red graph), electron fraction $Y_e$ (blue graph), temperature $T$
  (green graph), and velocity magnitude (magenta graph) are shown as a
  function of radius. The shock front (gray line labeled with
  $r_\mathrm{sh}$) can be seen in the discontinuities in density,
  temperature, and velocities at $r=168\,\mathrm{km}$.}
\label{fig:EOScomp_rhoYeT}
\end{figure}
In Figure~\ref{fig:EOScomp_rhoYeT}, we show the static fluid background
that comes from a spherically symmetric neutrino radiation
hydrodynamics simulation of the collapse of a $11.2M_\odot$ star
\citep{woosley:02} at $100\,\mathrm{ms}$ after core bounce using the
NSY code \citep{nagakura:17a}. For the calculations in this paper, the
NSY code is again used to calculate a steady-state solution of the
neutrino radiation field on this background using the DO method. The
opacities and emissivities are then exported to {\tt Sedonu}, which
computes a steady-state radiation field on the same
background. 

\begin{figure}
\includegraphics[width=\linewidth]{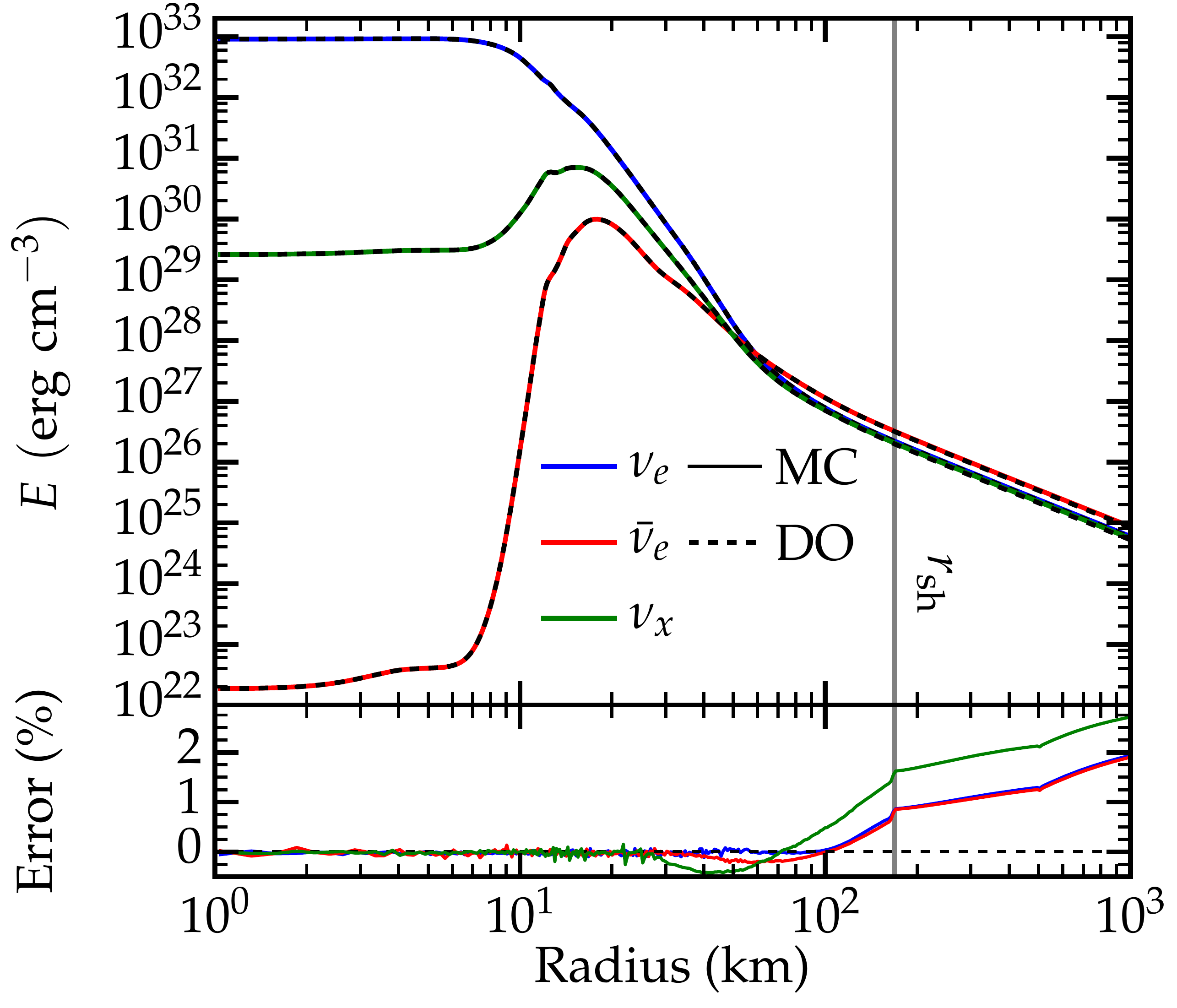}
\caption{\textbf{1D Neutrino Energy Density.} The total lab-frame
  neutrino energy density as a function of radius using the 1D\_4x DO
  calculation and the 1D\_4x\_native MC calculation. The error is
  defined as $(E_\mathrm{MC}-E_\mathrm{DO}) /
  (E_\mathrm{MC}+E_\mathrm{DO})$. There is excellent agreement between
  MC and DO inside the shock. The small differences between
  $30\,\mathrm{km}\lesssim r \lesssim 60\,\mathrm{km}$ are due to the
  error from the MC random walk approximation and
  decreases with MC random walk sphere size (see
  Appendix~\ref{app:montecarlo}). The differences between
  $70\,\mathrm{km}\lesssim r\leq r_\mathrm{sh}$ results from
  momentum-space diffusivity in the NSY code in strongly
  forward-peaked regions and improves with angular resolution. The
  differences outside of the shock come from slight non-conservation
  experienced in the NSY code due to finite spatial resolution.}
\label{fig:1Dedens}
\end{figure}
In Figure~\ref{fig:1Dedens} we show radial profiles of the total
energy density of each neutrino species using both the 1D\_4x DO
calculation and the 1D\_4x\_native MC calculation. For each species,
the energy density is approximately constant in the inner core
($r\lesssim 10\,\mathrm{km}$) as neutrinos are trapped and in
equilibrium with the fluid. In this region, the neutrinos have a
Fermi-Dirac distribution function, and so the energy density is
determined entirely by the fluid temperature and the electron and
nucleon chemical potentials. In the outer (i.e., shock-processed) core
at $r\gtrsim10\,\mathrm{km}$, the temperature is very high
($T\approx20\,\mathrm{MeV}$), and many more electron anti-neutrinos
and heavy lepton neutrinos are emitted than in the cooler inner
core. Beyond the energy-averaged neutrinospheres
($30\,\mathrm{km}\lesssim r \lesssim 60\,\mathrm{km}$, depending on
the neutrino species), the neutrinos are only weakly coupled to the
fluid and the energy density decreases as $E\propto r^{-2}$.

Both codes produce remarkably similar results, with differences in the
energy density (Figure~\ref{fig:1Dedens}) smaller than 2\% everywhere
within the shock. The remaining differences near the energy-averaged
neutrinospheres ($30\,\mathrm{km}\lesssim r \lesssim 60\,\mathrm{km}$,
depending on species) are due to the MC random walk approximation, and
decrease when the critical optical depth is increased (see
Appendix~\ref{app:montecarlo}). The differences outside of the
decoupling region ($r\gtrsim 40-80\,\mathrm{km}$, depending on
species) decreases with increasing DO directional angular
resolution. Outside of the shock, the difference in the energy density
grows as the NSY code experiences a small departure from $r^{-2}$
scaling of the energy density. This is an artifact of the finite
spatial resolution. The size of the error visibly increases at
$500\,\mathrm{km}$, where the radial resolution coarsens.

\begin{figure}
  \includegraphics[width=\linewidth]{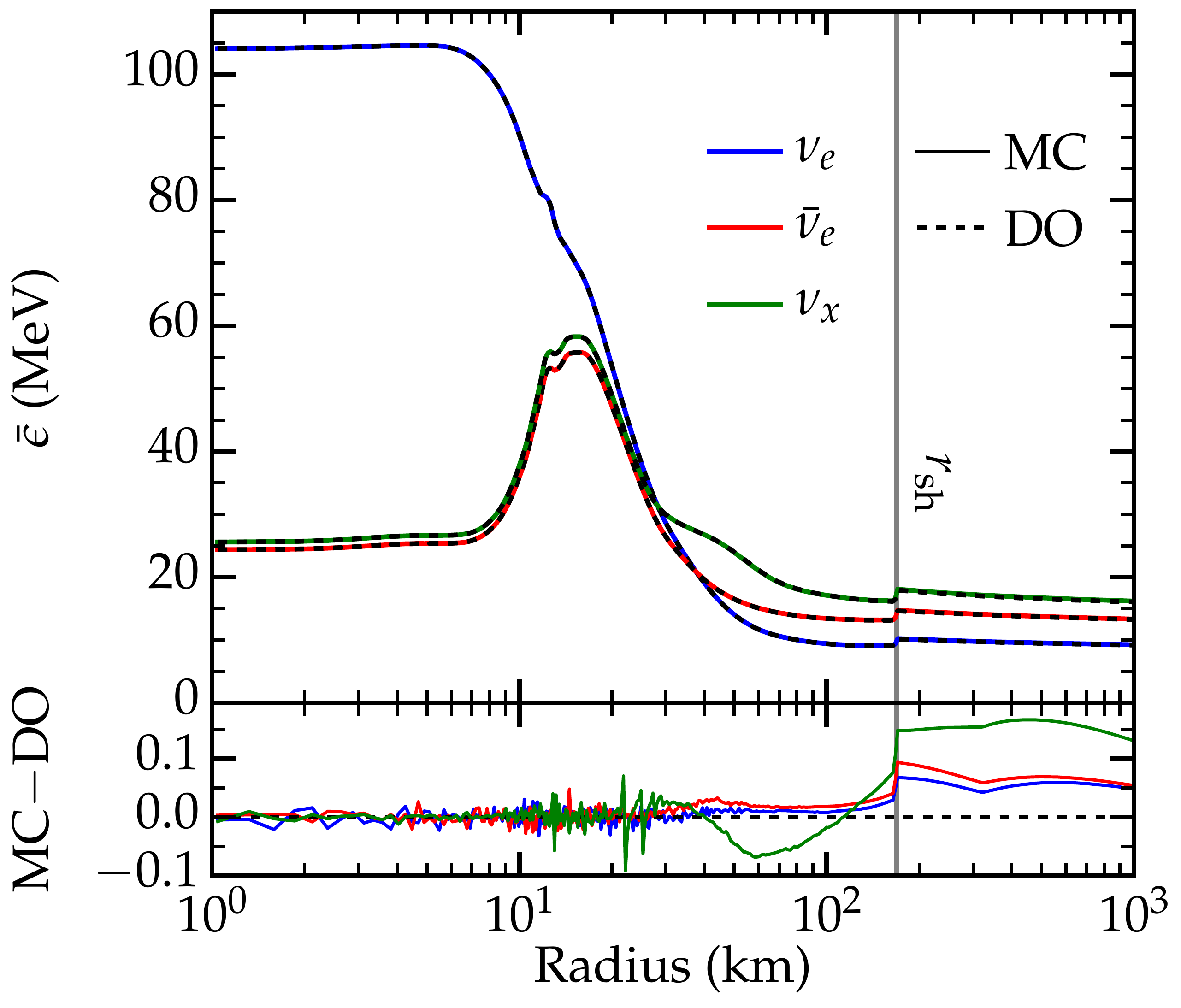}
  \caption{\textbf{1D Neutrino Average Energy.} The average
    comoving-frame neutrino energy weighted by the lab-frame energy
    density (Equation~\ref{eq:avge}) for all three neutrino species
    using the 1D\_4x DO calculation and the 1D\_4x\_native MC
    calculation. There is good agreement between MC and DO. Resolution
    tests show that the error improves with momentum-space resolution and MC
    particle random walk sphere size (see
    Appendix~\ref{app:montecarlo}). The average energy jumps at the
    shock, since the neutrino energy is blueshifted in the comoving
    frame in the supersonically infalling material outside the shock.}
  \label{fig:1DavgE}
\end{figure}
Figure \ref{fig:1DavgE} shows a quantity akin to the comoving-frame
average neutrino energy\footnote{Recall that we used mixed frame
  quantities, since many GR transport schemes are formulated in terms
  of lab-frame energy density and comoving-frame neutrino energy
  (e.g., \citealt{shibata:11}).}, defined as
\begin{equation}
  \label{eq:avge}
  \bar{\epsilon} = \frac{\sum E_i}{\sum E_i/\epsilon_i}\,\,,
\end{equation}
where $E_i$ is the lab-frame energy density in bin $i$ and
$\epsilon_i$ is the comoving-frame bin central energy. Because
neutrinos are in equilibrium with the fluid below the decoupling
region ($r=30-70\,\mathrm{km}$, depending on species), they have a
Fermi-Dirac distribution function that depends only on the fluid
temperature and electron and nucleon chemical potentials. The neutrino
absorption opacity scales approximately as $\kappa_a \sim \epsilon^2$,
so high-energy neutrinos are preferentially absorbed, causing the
average neutrino energy to continuously decrease with radius. The
average energy jumps at the shock front, since after passing the shock
front, neutrinos are moving through matter falling with speeds of
$|v|\sim0.1c$. The comoving-frame neutrino energy is thus Doppler
boosted even though the lab-frame energy density is constant across
the shock.

The differences in the average neutrino energy between the 1D\_4x DO
calculation and the 1D\_4x\_native MC calculation are smaller than
about $0.1\,\mathrm{MeV}$ within the shock. Analyzing the various
potential sources of errors, the differences at
$r\lesssim30\,\mathrm{km}$ are simply statistical noise that decreases
with the square root of the number of MC neutrino packets
simulated. The differences below the shock are primarily due to the MC
random walk approximation error, and decreases with an increased
critical optical depth (see Appendix~\ref{app:montecarlo}) independent
of the momentum-space resolution. The differences near and outside the
shock are a result of finite energy resolution, which results in
interpolation error when the NSY code transforms energy and direction
bins between the comoving and lab frames.

\begin{figure}
  \includegraphics[width=\linewidth]{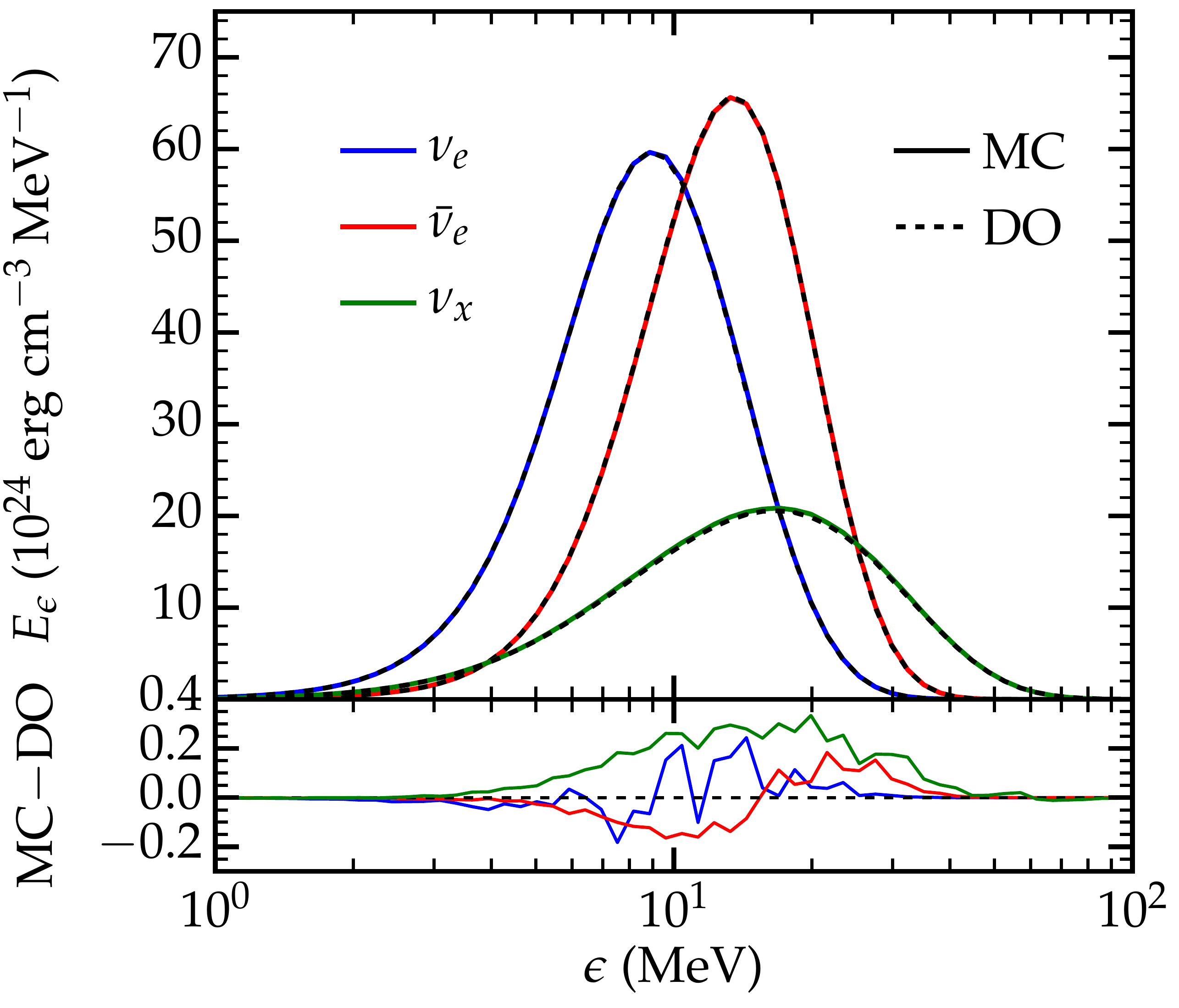}
  \caption{\textbf{1D Spectrum}. The lab-frame direction-integrated
    spectral energy density based on the comoving-frame neutrino
    energy density
    ($E_\epsilon=dE_\mathrm{lab}/d\epsilon_\mathrm{com}$) at
    $r=105\,\mathrm{km}$ for each species in the 1D\_4x DO calculation
    and the 1D\_4x\_native MC calculation. This point is inside the
    shock in the semi-transparent region. In the bottom panel are the
    differences between the MC and DO results, in the same
    units. There is good agreement between MC and DO that improves
    with DO and MC energy resolution. The error below
    $2\,\mathrm{MeV}$ is due to the large width ($2\,\mathrm{MeV}$) of
    the first energy bin. The oscillating errors above 10 MeV are
    artifacts from the two-grid DO method used in the NSY code
    (Appendix~\ref{app:boltzmann}, \citealt{nagakura:14}). Resolution
    tests show that the agreement improves with DO and MC
    momentum-space resolution. Both sources of error disappear in
    nonrelativistic calculations.}
  \label{fig:1Dspectra}
\end{figure}
We plot the direction-integrated neutrino energy density spectra at
$r=105\,\mathrm{km}$ for each neutrino species in
Figure~\ref{fig:1Dspectra}. This point is below the shock in the
semi-transparent region. The results of the 1D\_4x DO calculation and
the 1D\_4x\_native MC calculations agree in every bin with at most
$\approx1.5\%$ of the peak value. In energy bins with little energy
density, the relative errors become quite large, but bins with such
small energy density have much less dynamical effect in CCSN
simulations. Some statistical noise from the MC calculation can be
seen in the range of $5-20\,\mathrm{MeV}$, but the small overall
offsets are due to the finite neutrino energy and angular resolution.

\begin{figure}
  \includegraphics[width=\linewidth]{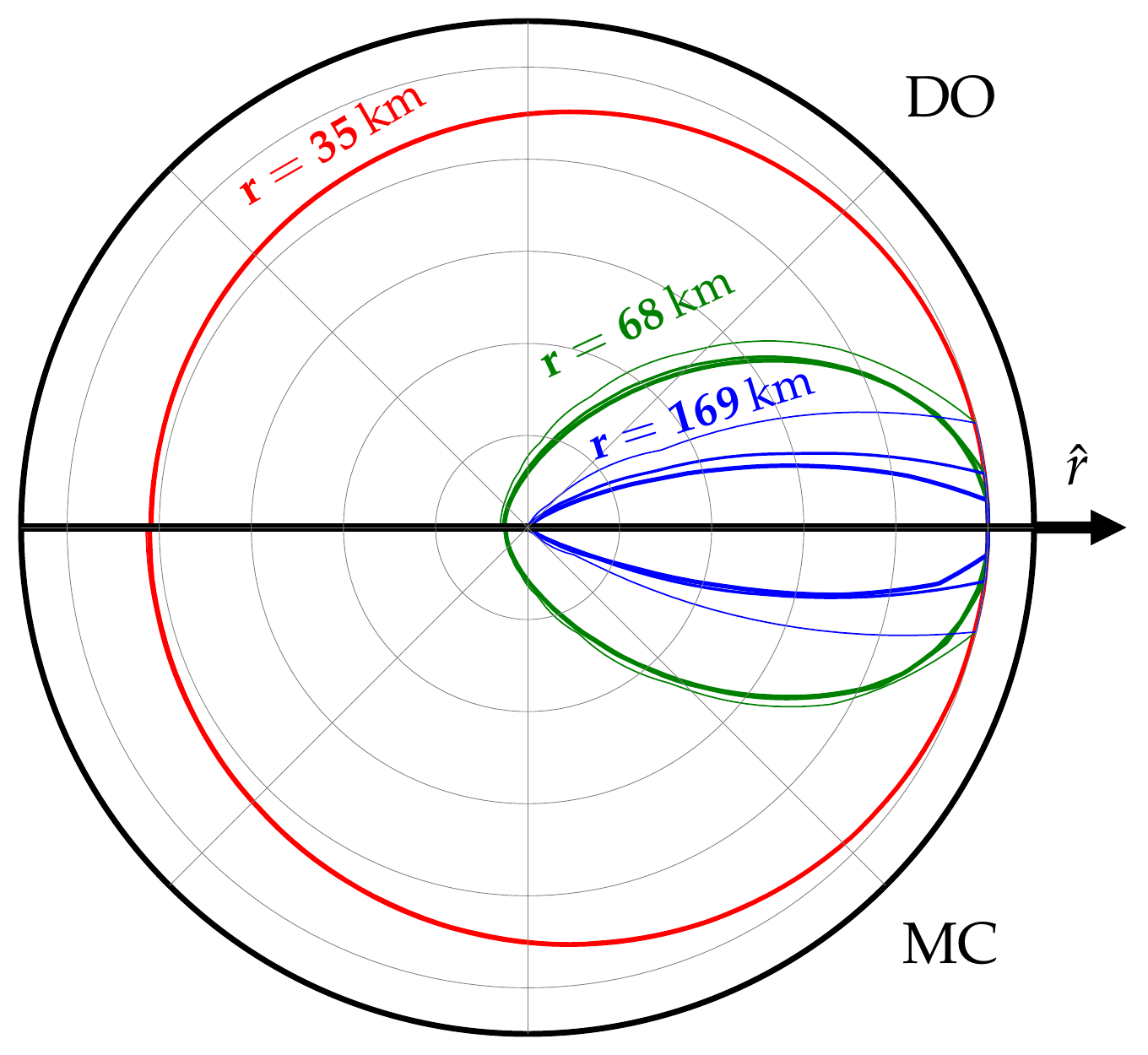}
  \caption{\textbf{1D Distribution Function Shape}. Normalized
    lab-frame distribution functions as a function of propagation
    angle for heavy lepton neutrinos at three radii using both the DO
    (top) and MC (bottom) calculations. A circular shape indicates
    isotropic radiation, while sharper shapes indicate radiation
    moving primarily in one direction. The outward radial direction is
    to the right in the plot. The distribution functions at three
    locations are shown: just inside the the shock ($169\,\mathrm{km}$
    -- blue), at $68\,\mathrm{km}$ (green), and at
    $\rho=2\times10^{12}\,\mathrm{g\,cm}^{-3}$
    ($35\,\mathrm{km}$). Line thickness corresponds to resolution. The
    thickest lines are from the 1D\_4x MC and DO calculations, the
    medium lines are from 1D\_2x MC and DO calculations, and the
    thinnest lines are from the 1D\_1x MC and DO calculations. The
    1D\_4x\_native calculation does not use an angular grid and so
    cannot be plotted here. The MC and DO results are nearly
    indistinguishable. The resolution does not affect nearly isotropic
    radiation fields, but low resolution artificially broadens
    free-streaming neutrino distributions.}
  \label{fig:1Dwireframe}
\end{figure}
Figure~\ref{fig:1Dwireframe} shows the lab-frame energy-integrated
heavy lepton neutrino distribution function at three separate
radii. The red curves are from the radius where
$\rho=2\times10^{12}\,\mathrm{g\,cm}^{-3}$ and show that the
neutrinos are nearly isotropic. The blue curves are from near the
shock front and are nearly free-streaming and very forward-peaked, as
almost no neutrinos are moving inward. The green lines show the
distribution function at an intermediate location of
$r=68\,\mathrm{km}$ ($\rho=4.7\times10^{10}\,\mathrm{g\,cm}^{-3}$)
between the trapped and free-streaming limits. In the plot, the
distribution function is normalized by the largest value so the shapes
can be easily compared. We assume a constant value for the
distribution within the directional bin in the forward direction and
linearly interpolate the distribution function for all other
directions. This is done to ensure that in post-processing the value
of the distribution function never exceeds one. However, this gives
rise to the artificially flattened nose of the distribution functions
most apparent in the blue curves.

The thickest lines in Figure~\ref{fig:1Dwireframe} are from
high-resolution 1D\_4x MC and DO simulations, while thinner lines
indicate lower resolution. The 1D\_4x\_native MC simulation does not
collect data on a grid of discrete angular bins, so its results cannot
be used to make such a plot. The importance of the angular resolution
is very apparent for the blue curves at the shock front, since most of
the neutrino energy is in a single angular bin. The MC results look
remarkably similar to the DO results, though a lack of numerical
diffusion in the MC calculations allows for slightly more sharply
forward-peaked distribution functions for a given angular resolution.
This angular dependence is reflected in all angular moments of the
distribution function.

\begin{figure}
  \includegraphics[width=\linewidth]{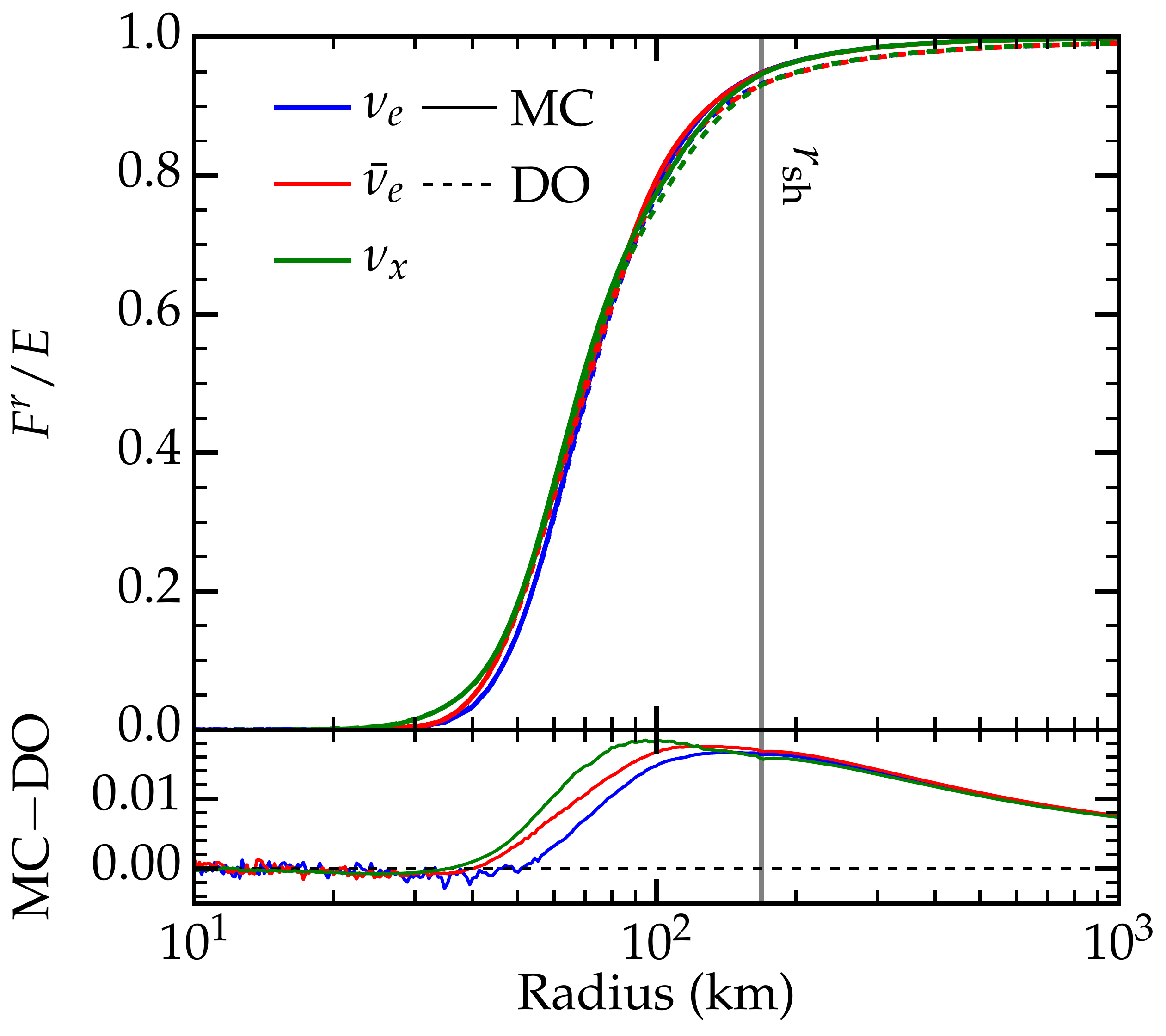}
  \caption{\textbf{1D Flux Factor}. The energy-integrated flux factor
    ($F^r/E$) for all three neutrino species in the lab frame using
    the 1D\_4x DO calculation and the 1D\_4x\_native MC
    calculation. The bottom panel contains the differences between the
    MC and DO results in the same units. There is good agreement
    between the MC and DO results that improves with DO directional
    angular resolution. The largest differences are in the
    semi-transparent region, where momentum-space diffusivity in the DO
    method broadens the distribution function angular shape.}
  \label{fig:1DFr}
\end{figure}
In Figure~\ref{fig:1DFr}, we show the energy-integrated lab-frame
radial flux factor ($F^r/E$, see Equation~\ref{eq:moments}) of the
distribution function of all three neutrino species using the 1D\_4x
DO calculation and the 1D\_4x\_native MC calculation. Below around
$30\,\mathrm{km}$ the neutrinos are trapped and the distribution
function is nearly isotropic, resulting in a minuscule flux relative
to the energy density (corresponding to the red curves in
Figure~\ref{fig:1Dwireframe}). In the transition region (corresponding
to the green curves in Figure~\ref{fig:1Dwireframe}), an increasing
fraction of the neutrino radiation energy is moving radially outward,
causing the flux factor to approach 1 at large radii (corresponding to
the blue curves in Figure~\ref{fig:1Dwireframe}). $F^\theta$ and
$F^\phi$ are identically zero due to spherical symmetry.

The angular moments of the radiation field are naturally sensitive to
the angular grid resolution. We see small differences of at most
around 0.02 in the flux factor, but the size of this difference scales
approximately linearly with the angular grid size for calculations
with a coarser angular grid (not plotted). {\tt Sedonu} consistently
predicts a more rapid transition to free streaming than does the NSY
code. Here the MC method shows a significant advantage in that by
computing moments directly rather than post-processing from an angular
grid, we get angular moments with effectively infinite angular
resolution. The NSY code comes very close to this solution, but
suffers from some angular diffusion. This causes the NSY code to
predict distribution functions that are slightly, though artificially,
more isotropic. The difference approaches a small but constant value
at large radii, where almost all of the energy in the DO calculations
is contained in the single outward-pointing angular bin. The {\tt
  Sedonu} results are, however, visibly noisy in the difference plot,
since subtractive cancellation tends to amplify statistical
noise. There is a small hump visible in the heavy lepton neutrino
difference plot just below $r=100\,\mathrm{km}$ that is a result of
the MC random walk approximation. The size of this hump decreases when
the critical optical depth is increased (see
Appendix~\ref{app:montecarlo}), bringing it closer to the electron
neutrino and electron anti-neutrino difference curves.

\begin{figure}
  \includegraphics[width=\linewidth]{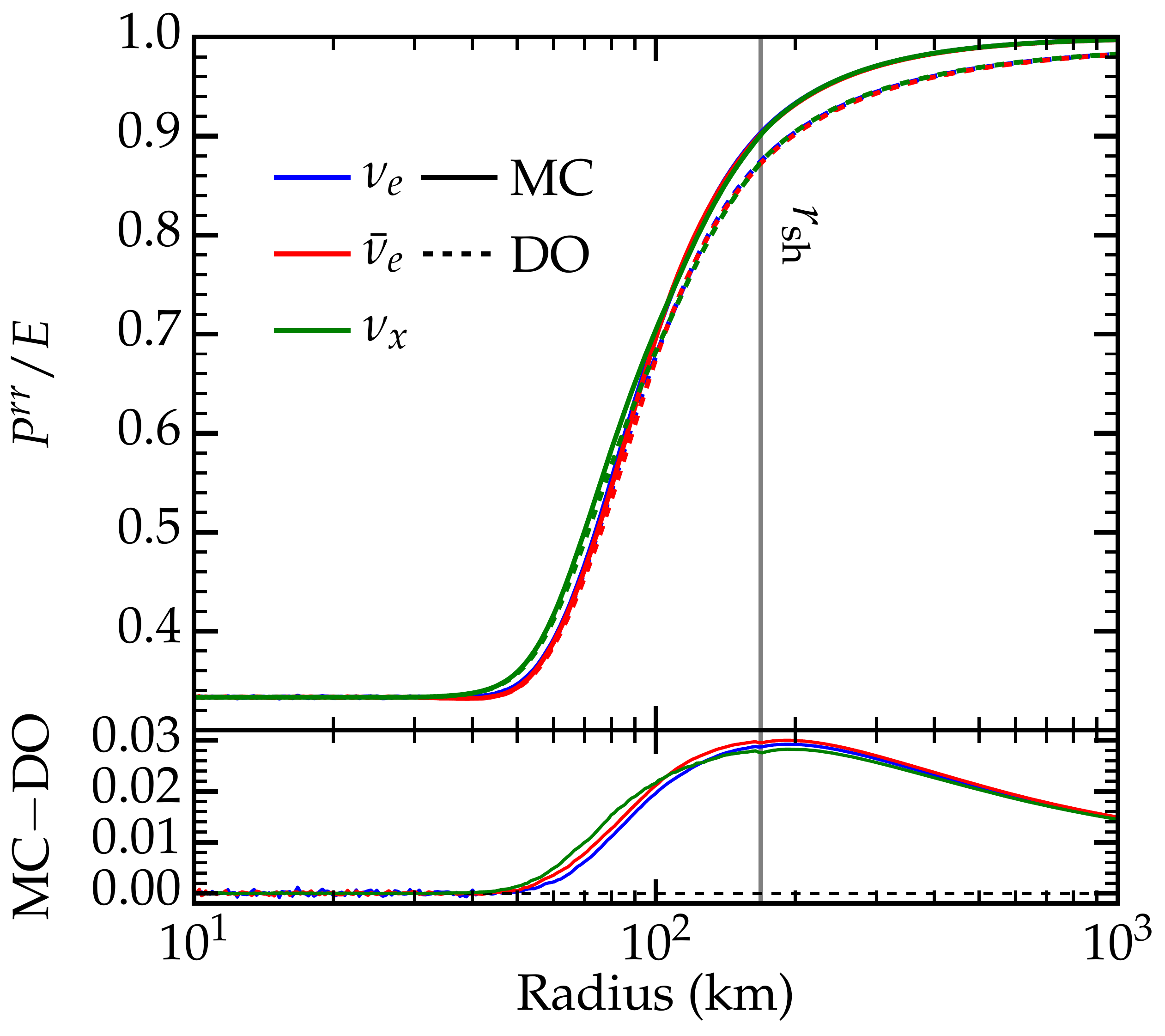}
  \caption{\textbf{1D Eddington Factor}. The energy-integrated $rr$
    component of the Eddington tensor ($P^{rr}/E$) in the lab frame
    for all three neutrino species using the 1D\_4x DO calculation and
    the 1D\_4x\_native MC calculation. The bottom panel shows the
    differences between the MC and DO results in the same units. There
    is good agreement between MC and DO that improves with DO
    directional angular resolution (see also
    Figure~\ref{fig:1Dresolution}). MC predicts a faster transition to
    free streaming.}
  \label{fig:1DPrr}
\end{figure}
In Figure~\ref{fig:1DPrr}, we show the $rr$ component of the
energy-integrated lab-frame Eddington tensor ($P^{ij}/E$, see
Equation~\ref{eq:moments}) of the distribution function of all three
neutrino species. Only the diagonal components of the Eddington tensor
($P^{rr}/E$,$P^{\theta\theta}/E$, and $P^{\phi\phi}/E$) are nonzero in
spherical symmetry. At $r\lesssim30\,\mathrm{km}$, all diagonal
components of the Eddington tensor are $1/3$ because the radiation is
nearly isotropic. After decoupling, the $rr$ component approaches
unity as all radiation is moving radially outward, while the
$\theta\theta$ and $\phi\phi$ components (not plotted) approach zero.

Once again, the differences between {\tt Sedonu} and the NSY code
scale approximately linearly with the neutrino direction angular zone
sizes.  However, the maximum difference of 0.03 is larger than the
maximum flux factor difference of 0.02. Unlike with previously discussed
radiation quantities, the random walk approximation does not add
significant error to $P^{rr}$. Though we do not plot
$P^{\theta\theta}$ or $P^{\phi\phi}$, the differences between MC and
DO results are similar to those of $P^{rr}$. Since the integral in
Equation~\ref{eq:moments} contains a factor of $(\mathbf{\Omega}\cdot
\hat{r})^2$, the results do not suffer from subtractive cancellation
and the amount of statistical noise is significantly lower than that
of the flux factor.

\begin{figure}
  \includegraphics[width=\linewidth]{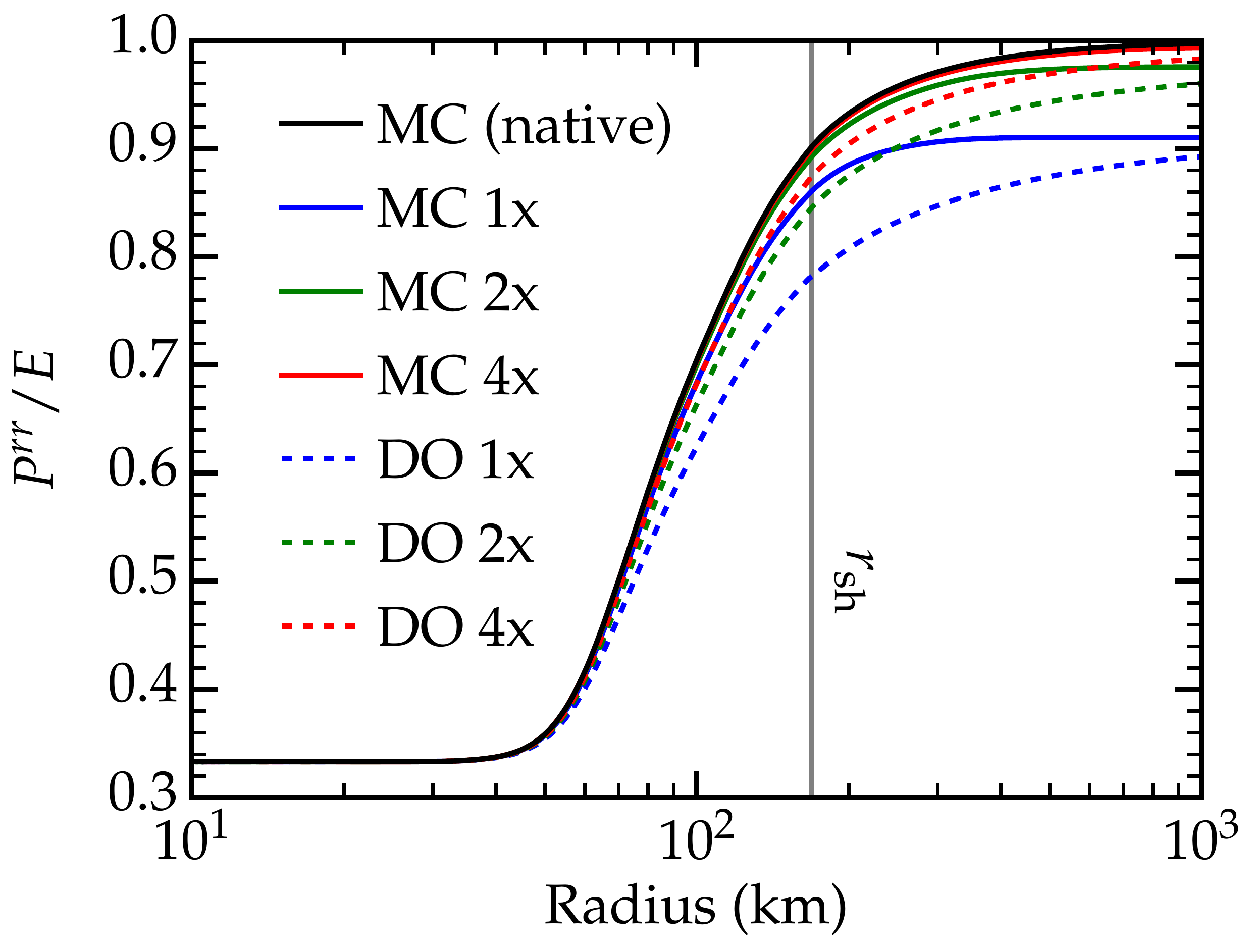}
  \caption{\textbf{1D Resolution}. The $rr$ component of the lab-frame
    energy-integrated Eddington tensor as calculated using the DO
    transport method (dashed lines) and the MC method (solid
    lines). The black solid line shows the results from the
    1D\_4x\_native calculation (MC particles accumulate into moments
    directly). The colored solid lines show results from MC
    calculations where MC particles collect into angular bins (1D\_1x,
    1D\_2x, and 1D\_3x for blue, green, and red curves, respectively),
    which are post-processed in the same manner as the DO
    results. Very high directional angular resolution is required for
    accurate angular moment results.}
  \label{fig:1Dresolution}
\end{figure}

The dependence of the angular moments on angular resolution can be
clearly seen in Figure~\ref{fig:1Dresolution}, where we plot the $rr$
component of the Eddington tensor for four MC calculations (solid
lines) and three DO calculations (dashed lines). We first direct our
attention to the blue lines, corresponding to the low resolution
1D\_1x DO and MC calculations. Even though both are post-processed in
the same way from the same angular grid, the MC results transition to
large values of $P^{rr}$ faster than do the DO results. At
$r\approx300\,\mathrm{km}$, $P^{rr}/E$ saturates at the maximum value
possible given the angular resolution, which the DO results approach
at large radii. The same is true for the higher resolution green and
red curves, but the saturation occurs at a larger radius and is not so
visibly obvious.

Due to the effectively infinite angular resolution of the
1D\_4x\_native MC calculation, the corresponding black line in
Figure~\ref{fig:1Dresolution} can be thought of as exact, modulo a
small amount of MC noise. Going from coarsest to finest resolution,
the maximum difference between the DO results and the black curve are
0.125, 0.057, and 0.028. This corresponds to a factor of 2.2
improvement when going from 1x to 2x resolution, and a factor of 2.0
when going from 2x to 4x resolution. Similarly, the maximum difference
between the gridded MC and the native MC results are, in order of
increasing resolution, 0.0896, 0.0243, and 0.0062. The accuracy
improves by a factor of 3.7 when going from 1x to 2x resolution, and
by a factor of 3.9 when going from 2x to 4x resolution. This trend,
where DO results are near first order convergence and gridded MC
results are near second order convergence, is apparent in the flux
factor results as well. This is because the post-processing angular
integration scheme is second order (except in the forward-most bin,
where it is first order), but the evolution scheme in the NSY code is
only first order.

\begin{figure}
  \includegraphics[width=\linewidth]{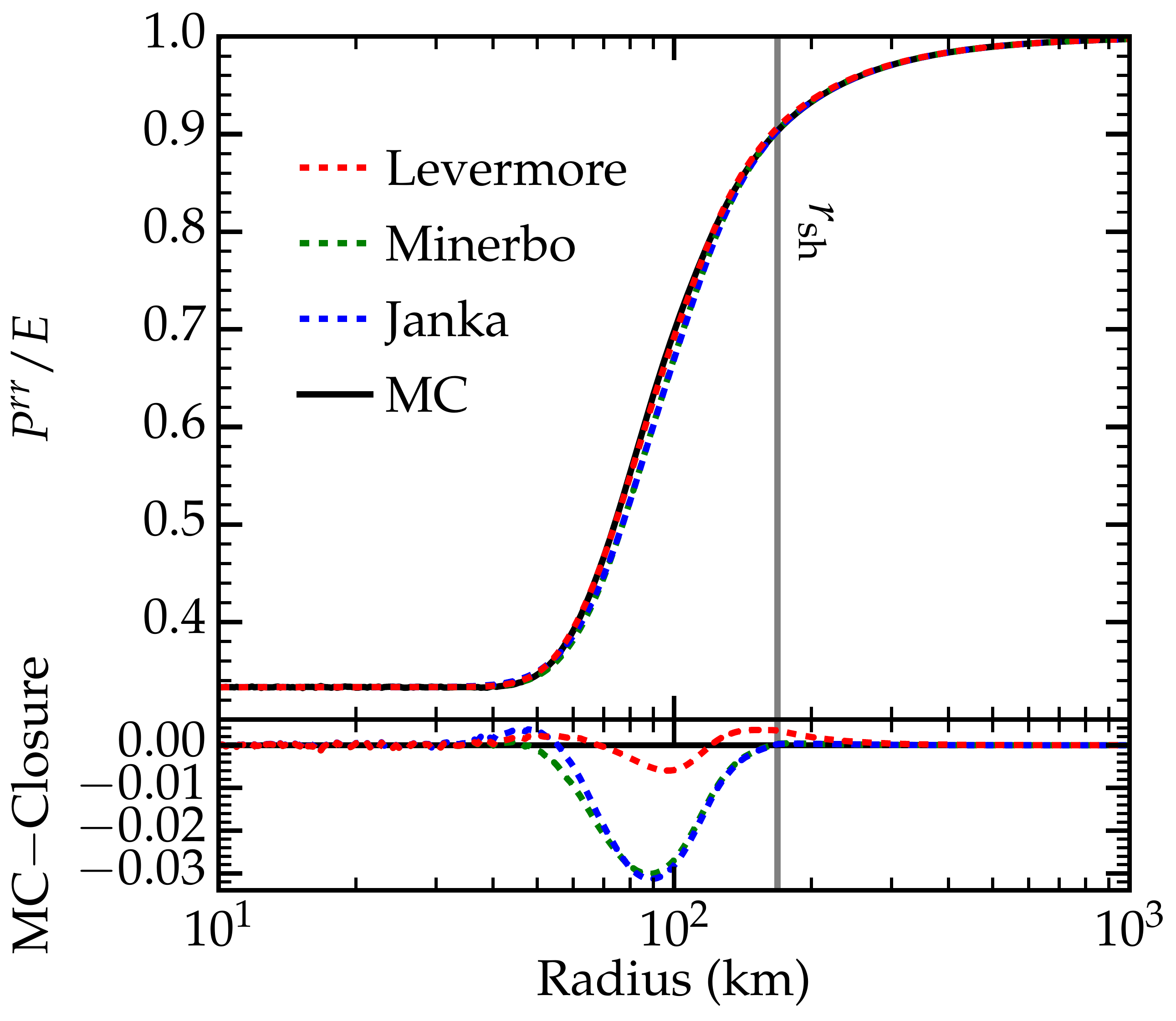}
  \caption{\textbf{1D Approximate Closures}. The lab-frame $rr$
    components of the energy-integrated Eddington tensor as calculated
    in the 1D\_4x\_native MC calculation (black solid line), the
    Levermore closure (red dashed lines), the Minerbo closure (green
    dot-dashed lines), and the Janka closure (blue dotted lines). The
    approximate closure values are calculated using the energy density
    and flux from the 1D\_4x\_native MC calculation. The bottom panel
    shows the difference between the MC results and the closure
    estimate in the same units. The Levermore closure appears to have
    the closest agreement with the MC results in this scenario. The
    differences between the approximate closures and highest
    resolution DO results (not plotted) are nearly identical to the
    differences between the approximate closures and MC results. The
    errors in the $\theta\theta$ component of the Eddington tensor
    (not plotted) behave in the same way.}
  \label{fig:1Dclosures}
\end{figure}
The use of an approximate analytic closure in two-moment radiation
transport schemes is significantly faster than either the DO or MC
methods. However, since there are many reasonable closures available,
it is of great interest to evaluate how well these closures perform
against our full Boltzmann results. We re-plot the electron neutrino
$P^{rr}$ curve (black line in Figure~\ref{fig:1Dresolution}) from the
1D\_4x\_native MC calculation as a solid black line in
Figure~\ref{fig:1Dclosures}. We then use $E$ and $F^r$ from the same
MC calculation to estimate $P^{rr}$ using the three analytic closures
given in Equation~\ref{eq:closures}. The Janka and Minerbo closures
perform similarly and have a maximum difference with MC of $\sim0.03$,
which is comparable to the differences between the 1D\_4x DO
calculation and the 1D\_4x\_native MC calculation. The Levermore
closure, however, performs better, with a maximum difference of
$\sim0.006$. This is significantly better than the accuracy of any DO
result and is comparable to the accuracy of the 1D\_4x MC
calculation. These results are also replicated in a similar analysis
of $P^{\theta\theta}$ and $P^{\phi\phi}$ (not plotted). In short,
analytic closures perform remarkably well in this particular
steady-state spherically-symmetric transport calculation.

\begin{figure}
  \includegraphics[width=\linewidth]{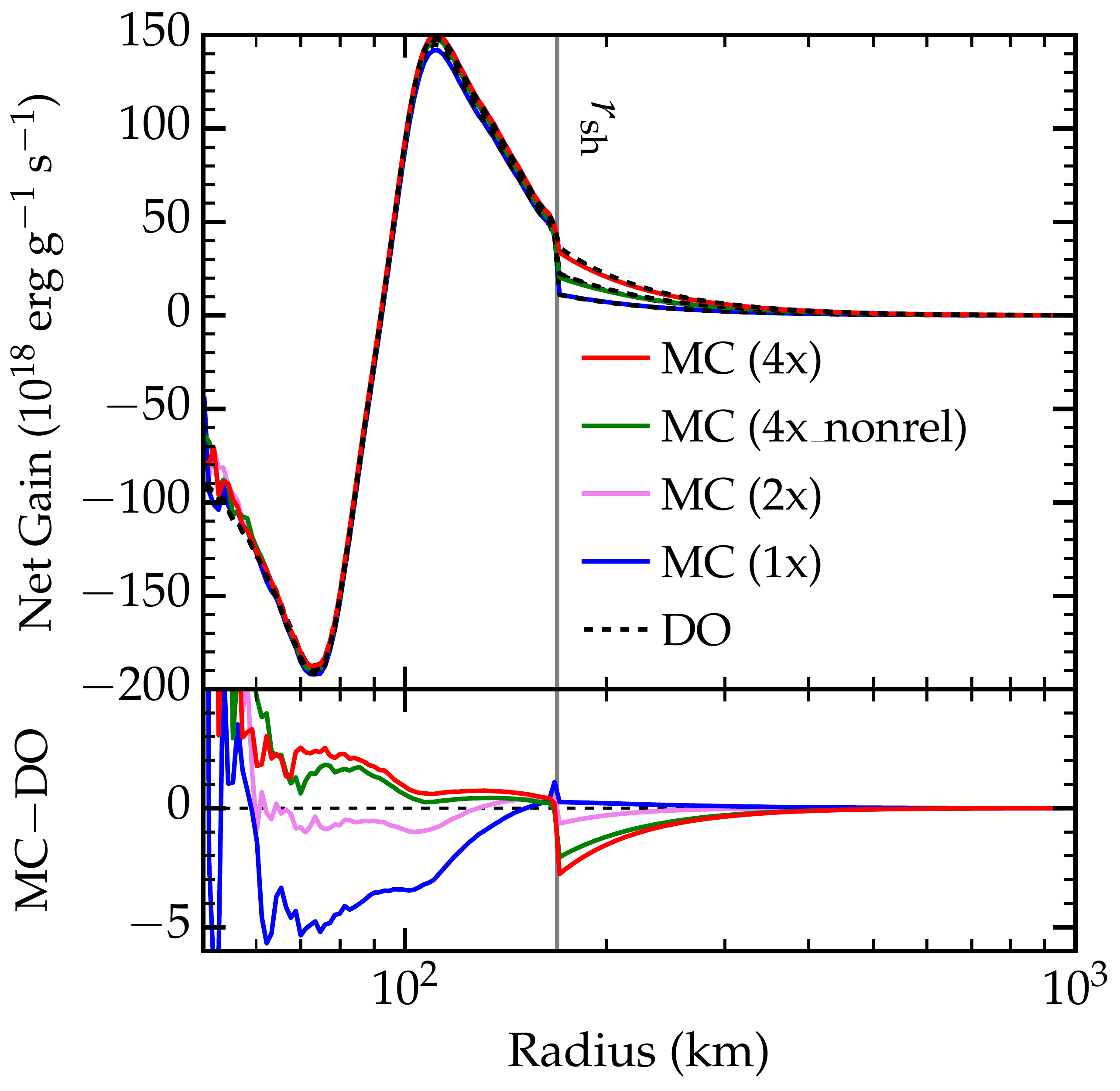}
  \caption{\textbf{1D Net Gain}. The net gain (heating $-$ cooling)
    using the 1D\_4x DO calculation and the 1D\_4x\_native MC
    calculation. The transition from net cooling to net heating lies
    at $\sim90\,\mathrm{km}$. The MC results are very noisy below
    $\sim 60\,\mathrm{km}$, but otherwise the highest resolution MC
    and DO results agree within the shock to $\sim1\%$. Neutrino pair
    annihilation is the dominant heating mechanism outside the
    shock. This process is under-resolved in neutrino energy space,
    but tests show this heating rate to converge with $\geq 160$
    energy bins. The jump in the difference at the shock is due to the
    large jump in density combined with an over-estimate of the
    heating rate from our neutrino pair annihilation treatment.}
  \label{fig:1Dgain}
\end{figure}
The primary role of neutrinos in the explosion mechanism of CCSNe is
redistributing thermal energy from the protoneutron star region to the
gain region that drives turbulence and pushes the shock outward. The
relevance of these detailed transport calculations comes down to how
the differences between the methods affect the heating and cooling of
matter in the supernova. In Figure~\ref{fig:1Dgain}, we show the
comoving-frame net gain, i.e., the heating rate less the cooling rate
due to neutrinos. We show results from the 1D\_4x MC and DO
calculations (red line), the 1D\_4x\_nonrel MC and DO calculations
(green line), and the 1D\_1x MC and DO calculations (blue line). Below
about $90\,\mathrm{km}$ the fluid is overall cooling, and the emitted
neutrinos pass through the gain region from $90\,\mathrm{km}$ to
$170\,\mathrm{km}$ and deposit energy. Below the shock, the net gain
from the 1D\_4x MC and DO calculations are very similar, with
differences of $\lesssim1\%$ of the peaks in the gain curve.

Just outside the shock, the fluid densities are low and most nucleons
are bound in nuclei. Because of this, the heating is due primarily to
neutrino pair annihilation. The pair annihilation rates are
under-resolved in neutrino energy space even with 80 energy bins,
resulting in significant differences between heating rates of
different resolutions. However, test results show only a $20\%$
difference between the 1D\_4x results and a test with an energy-space
resolution of 160 bins. We can use the radial profiles of heating
rate, density, and velocity to estimate the amount of energy per
nucleon the fluid is heated before passing through the shock as
\begin{equation}
  \Delta E \approx \int \frac{m_n \mathcal{H}(r)}{\rho(r) |v(r)|}dr\,\,,
\end{equation}
where $\mathcal{H}$ is the heating rate. Using the heating rate from
the highest-resolution simulations, this predicts a total heating of
$\sim0.1\,\mathrm{MeV\,nucleon}^{-1}$. Compared to the post-shock
temperatures of $T\lesssim20\,\mathrm{MeV}$ and the iron-56 binding
energy of $8.8\,\mathrm{MeV}$, this pre-shock heating is
insignificant.

The differences between MC and DO are amplified outside the shock,
where we must divide a volumetric heating rate (in erg cm$^{-3}$
s$^{-1}$) by the low density to get a specific heating rate. Also,
recall that in our calculation of pair annihilation rates, we assume
that the neutrino distribution functions are isotropic (see
Appendix~\ref{app:sourceterms} for details). At large radii relative
to the neutrinospheres and to leading order, however, the angular
dependence actual reaction rates is (e.g., \citealt{bruenn:85})
\begin{equation}
  R_\mathrm{pair,abs}(\epsilon,\bar{\epsilon},\mathbf{\Omega}\cdot\bar{\mathbf{\Omega}}) \propto 1-\mathbf{\Omega}\cdot\bar{\mathbf{\Omega}} \sim \left(\frac{r}{r_\nu}\right)^{-2}\,\,,
\end{equation}
where $r_\nu$ is the neutrinosphere radius. The location of the
neutrinosphere depends on the neutrino species and energy, but for a
typical radius of $r_\nu=50\,\mathrm{km}$ this angular term scales the
reaction rate by a factor of $\sim 0.1$ at the shock. Thus, we expect
the heating rates (and hence the heating rate differences) to be
over-estimated by a factor of $\sim10$ at the shock.

% vmax = 0.1285c
% gamma = 1.0084
% doppler = 1.138
Including velocity-dependence in neutrino transport algorithms is a
complication that can significantly increase the complexity and cost
of the transport calculation. It is natural to attempt to quantify the
size of the error made in codes that neglect velocity dependence. We
repeat the high-resolution calculations with the same rest-frame fluid
profile shown in Figure~\ref{fig:EOScomp_rhoYeT}, but set all
velocities to zero. Velocity dependence changes the comoving frame
neutrino energies and directions, modifying the rates at which
neutrinos interact with the fluid. This has a very minor effect below
the shock, but significantly changes the heating rate outside the
shock where velocities are $\sim 0.1c$. However, the density drops by
a factor of 10 across the shock and the pre-shock fluid moves so
quickly that the overall heating is negligible. These small errors
outside the shock are unlikely to have a significant impact on
simulation results. The volume-integrated net gain in the gain region
(where there is net heating under the shock) is
$2.16\times10^{51}\,\mathrm{erg\,s}^{-1}$ in the 1D\_4x DO calculation
and $2.18\times10^{51}\,\mathrm{erg\,s}^{-1}$ in the 1D\_4x\_native MC
calculation, a difference of only $0.34\%$. Compare this to the
difference of the same quantity between the 1D\_4x and 1D\_1x DO
calculations of $2.0\%$ and between the 1D\_4x and 1D\_4x\_nonrel DO
calculations of $1.2\%$. Though including velocity dependence impacts
the heating rates more than the differences between the codes in the
highest resolution case, low resolution can cause significantly larger
inaccuracies.

%%%%%%%%%%%%%%
% 2D RESULTS %
%%%%%%%%%%%%%%
\section{Transport Comparison in Axisymmetry}
\label{sec:2Dcomparison}

\begin{figure*}
  \centering
  \includegraphics[width=0.75\linewidth]{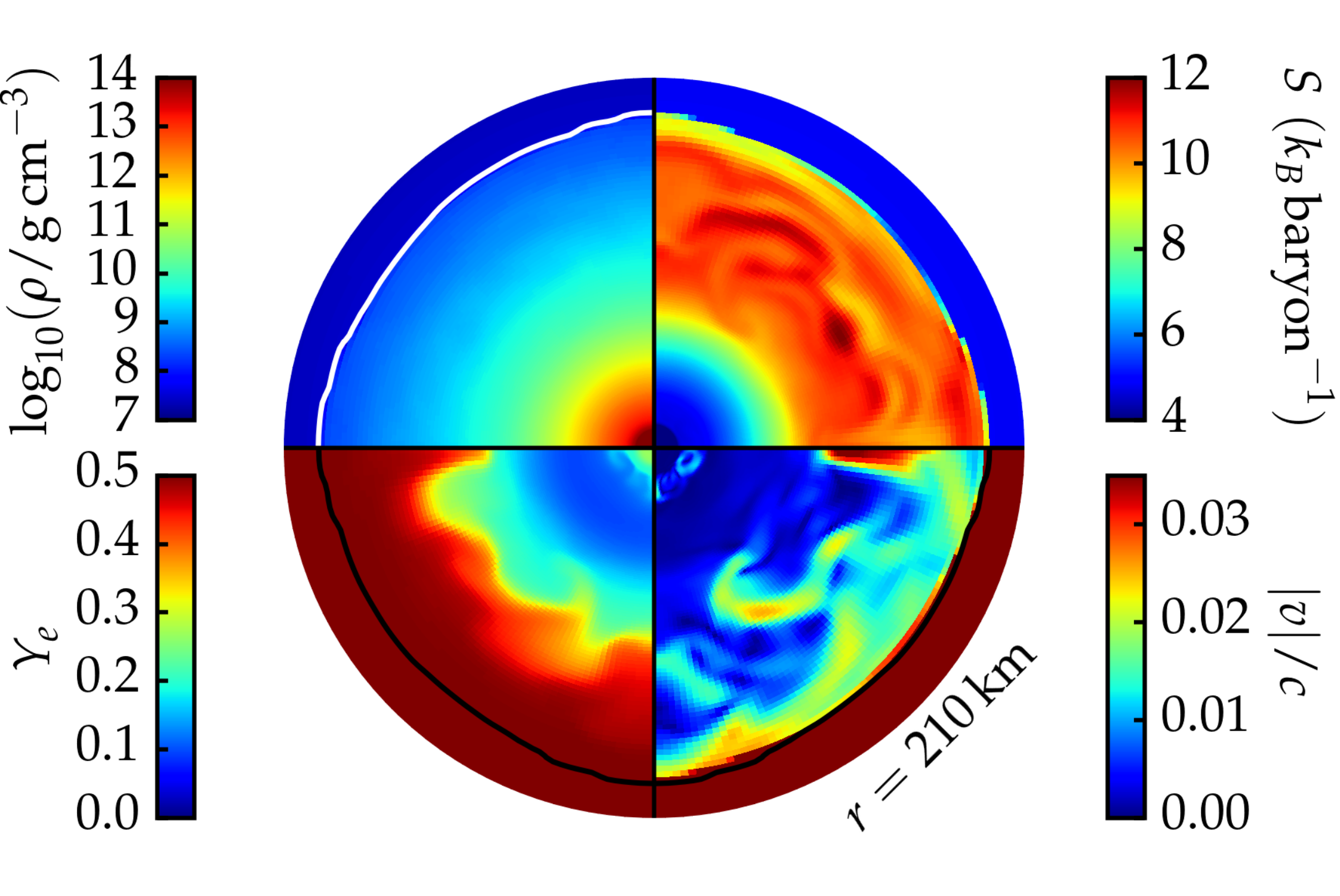}
  \caption{\textbf{2D Fluid Background.} The rest frame fluid density
    (top left), entropy (top right), electron fraction (bottom left),
    and speed (bottom right) from 2D core collapse simulations using
    the NSY code \citep{nagakura:17a} at $100\,\mathrm{ms}$ after
    bounce. The shock front is drawn as a contour at
    $S=7\,k_B\,\mathrm{baryon}^{-1}$ and is colored for clarity. The
    polar axis is vertical and the equatorial plane is horizontal. The
    gain region hosts neutrino-driven convection and protoneutron star
    convection is visible in the velocity and electron fraction
    quadrants. All quadrants in the plot show data from the northern
    hemisphere, though the computational domain includes both
    hemispheres. This fluid background is used for all axisymmetric
    simulations in this study.}
  \label{fig:2Dfluid}
\end{figure*}
In this section, we describe results from the first multi-dimensional
comparison of Boltzmann-level neutrino transport codes. We present a
set of four axisymmetric time-independent neutrino transport
calculations as listed in Table~\ref{tab:SimulationList}. Once again,
the NSY code is used to calculate an approximately steady-state
solution and the opacities and emissivities are exported from the
completed NSY calculations to \texttt{Sedonu} for the MC calculation. Due to
computational cost, we only consider two resolutions in the DO
code. The low-resolution (2D\_LR) calculations have momentum-space
resolution equivalent to the 1D\_1x calculations, and the
high-resolution (2D\_HR) calculations have momentum-space resolution
between that of the 1D\_1x and 1D\_2x calculations.

The rest-frame fluid profile used in all simulations shown in
Figure~\ref{fig:2Dfluid} comes from a 2D simulation of the collapse of
the same $11.2M_\odot$ star \citep{woosley:02} used in
Section~\ref{sec:1Dcomparison} \citep{nagakura:17a}. In
Figure~\ref{fig:2Dfluid} and in all other colormap plots in this
section, data separated into quadrants shows data from the northern
hemisphere of the calculation only to ease visual comparison of
datasets. Data on half-circles show the full simulation domain out to
$r=210\,\mathrm{km}$. In Figure~\ref{fig:2Dfluid}, multi-dimensional
structure in all fluid quantities is apparent and is due to
neutrino-driven turbulent convection. The postshock velocity field in
particular shows fluid speeds up to $0.037\,c$, compared to the
maximum radial velocity of $0.015\,c$ in the 1D calculations. This
multi-dimensional structure provides a challenge for any radiation
transport method.

\begin{figure*}
\includegraphics[width=\linewidth]{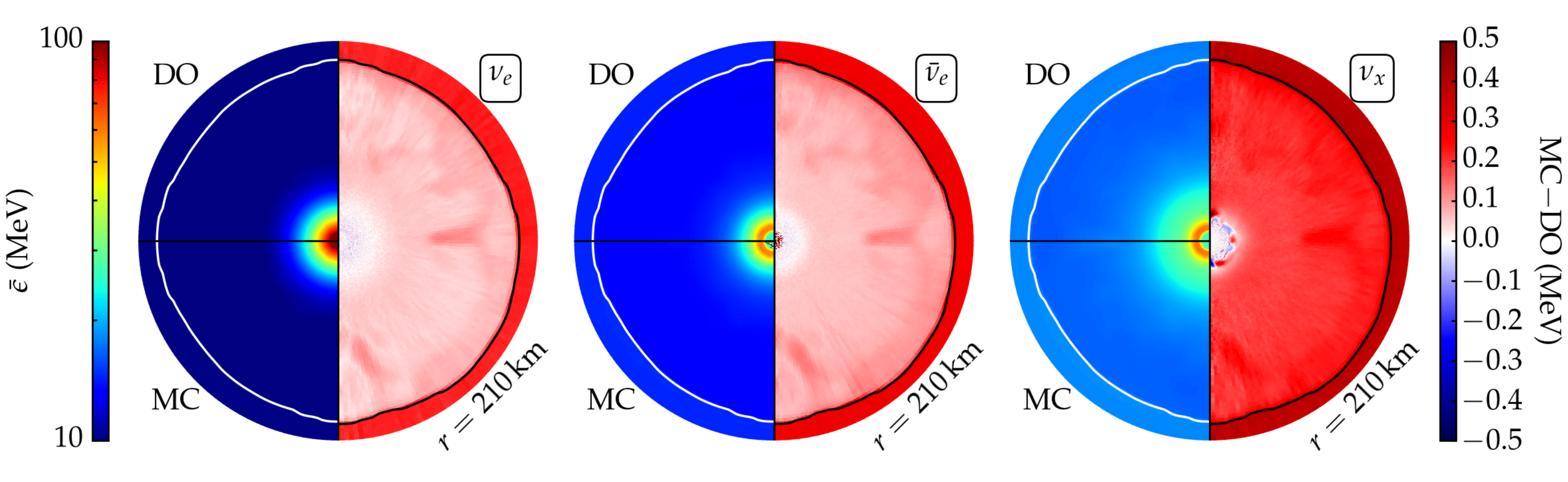}
\caption{\textbf{2D Neutrino Average Energy.} All plots show the
  comoving-frame average neutrino energy (Equation~\ref{eq:avge}). In
  each plot, we show results from the 1D\_HR DO calculation (top left
  quadrant, northern hemisphere data) and the 1D\_HR\_native MC
  calculation (bottom left quadrant, northern hemisphere data). The
  difference between them in MeV is shown in the right half of each plot,
  which contains data from both hemispheres. The left plot shows
  results from electron neutrinos, the center plot shows results from
  electron anti-neutrinos, and the right plot shows results from heavy
  lepton neutrinos. The shock front is drawn as a contour at entropy
  $S=7\,k_B\,\mathrm{baryon}^{-1}$ and is colored for clarity. The
  results agree well, though the error jumps across the shock due to
  diffusivity in the two-grid DO method with limited neutrino energy
  resolution.}
\label{fig:2DavgE}
\end{figure*}
We begin with a comparison of the spectral properties of the results
from {\tt Sedonu} and the NSY code. Figure~\ref{fig:2DavgE} shows the
comoving-frame average energy of each of the three simulated neutrino
species for the 2D\_HR DO calculation and the 2D\_HR\_native MC
calculation. Just as in the 1D calculations, the electron neutrinos
have the highest energy in the inner core due to the high electron
chemical potential, and the lowest energy at large radii since they
decouple at the largest radius and the lowest matter temperature. The
DO and MC results are nearly identical, so differences can only be
seen in the right half of each plot, where we subtract the DO results
from the MC results. The average energies differ between the MC and DO
results by at most 0.5 MeV, which is larger than in the 1D results in
Figure~\ref{fig:1DavgE} due to the lower energy resolution. When
electron neutrinos and anti-neutrinos decouple from matter, the
opacity is dominated by absorption. Because of this, MC packets that
use the random walk approximation quickly lose energy, preventing
errors from the random walk approximation from propagating outward
through the rest of the domain. However, the heavy lepton neutrino
opacity is dominated by scattering, so MC packets carrying errors from
the random walk approximation retain their energy when traveling
through the rest of the domain, causing errors to be slightly
higher. Just as in the 1D results, the differences between the MC and
DO average energies error jump across the shock, since the NSY code
suffers from numerical diffusion when transforming between grids in
the two grid approach (Appendix~\ref{app:boltzmann}). There are also a
number of hot spots in the average energy differences within the
shock, which correspond to regions of high velocity in
Figure~\ref{fig:2Dfluid}. These differences are also because of some
numerical diffusion in the two-grid approach. Heavy lepton neutrinos
are more strongly impacted by the protoneutron star convection, since
they decouple at a smaller radius. The features visible in the heavy
lepton neutrino average energy difference plot (rightmost panel of
Figure~\ref{fig:2DavgE}) at small radii are diminished by reducing the
MC random walk critical optical depth, independent of momentum space
resolution.

\begin{figure}
  \includegraphics[width=\linewidth]{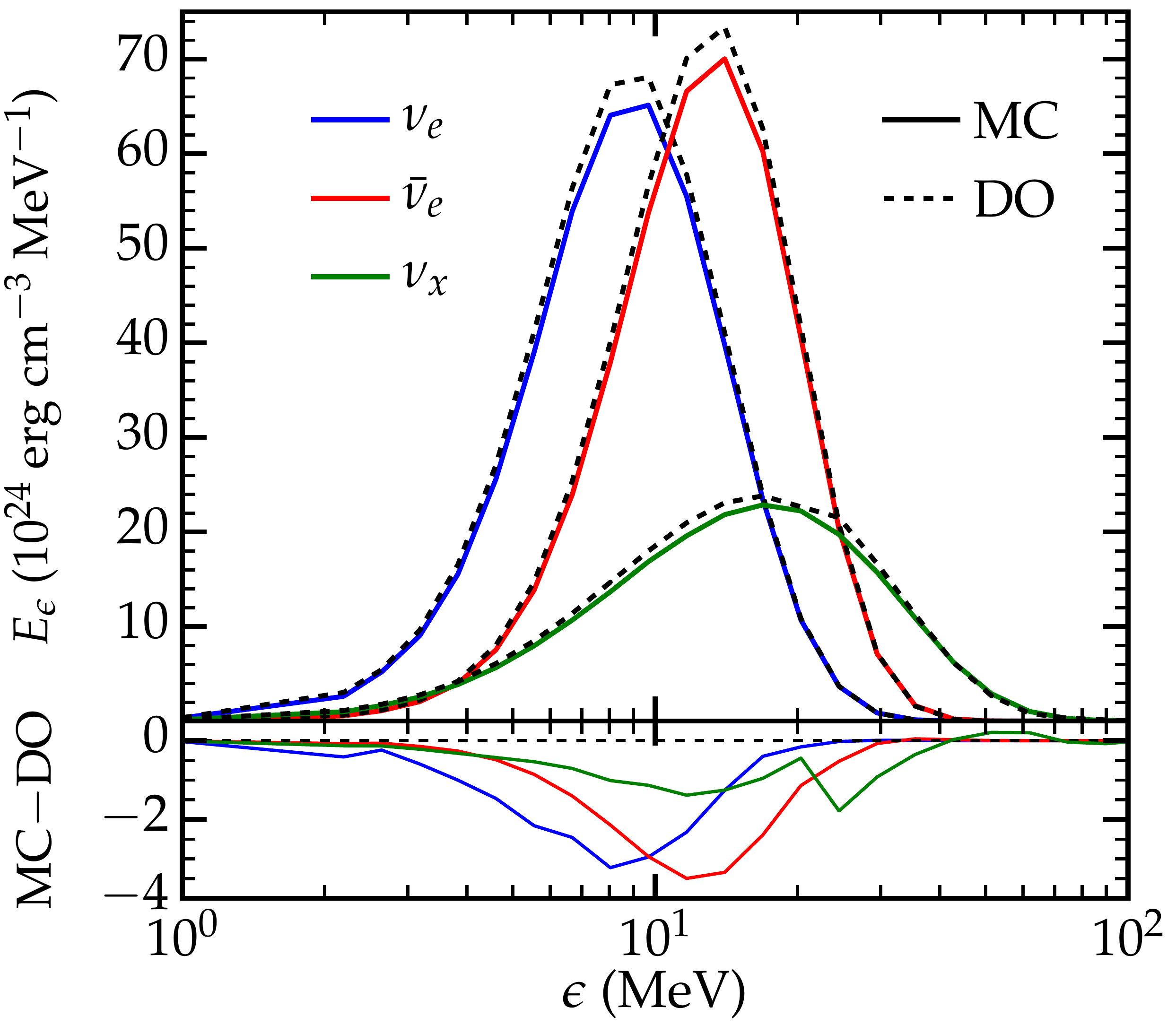}
  \caption{\textbf{2D Spectrum}. The lab-frame direction-integrated
    spectral energy density based on the comoving-frame neutrino
    energy density ($E_\epsilon =
    dE_\mathrm{lab}/d\epsilon_\mathrm{com}$) for each species at
    $r=105\,\mathrm{km}$ and $\theta=36^\circ$ (from the north
    pole). Dashed lines are results from the 2D\_HR DO calculation and
    solid lines are from the 2D\_HR\_native MC calculation. There is
    good agreement between the MC and DO results, though the DO
    results have lower peaks due to low angular resolution. The
    differences in the amplitudes reflects differences in overall
    energy density decrease with momentum-space resolution.}
  \label{fig:2Dspectrum}
\end{figure}
Though the energy resolution is coarser than in the 1D calculations,
we are able to compare the full spectra at a given location. For
Figure~\ref{fig:2Dspectrum}, we choose the same radius of
$105\,\mathrm{km}$ used for Figure~\ref{fig:1Dspectra} and an angle of
$36^\circ$ from the north pole. We plot the direction-integrated
spectra of all three neutrino species using the 2D\_HR DO calculation
and the 2D\_HR\_native MC calculation. The neutrino energy density
within each comoving frame energy bin is measured in the lab frame and
the individual neutrino energy in the comoving frame, resulting in a
mixed-frame quantity. The results are remarkably similar, and
effectively reproduce the 1D results. The heights of the peaks differ
by $\sim 5\%$, which is comparable to the differences between MC and
DO results in the energy density in the lower-resolution 1D
results.

\begin{figure*}
\includegraphics[width=\linewidth]{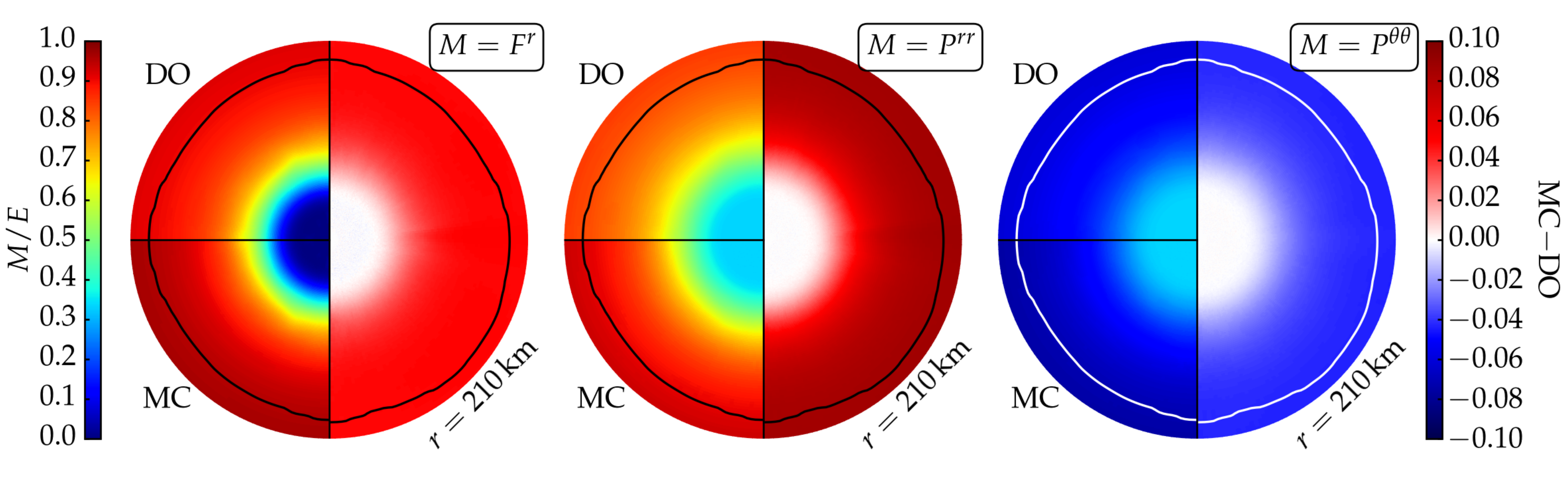}
\caption{\textbf{2D Flux Factor and Eddington Tensor (Diagonal).}  The
  leftmost plot shows the electron neutrino energy-integrated
  lab-frame radial flux factor, the center plot shows the $rr$
  component of the Eddington tensor, and the right plot shows the
  $\theta\theta$ component of the Eddington tensor. In each plot, we
  show the radiation moment computed using the 2D\_HR DO calculation
  (top left quadrant, northern hemisphere data) and the 2D\_HR\_native
  MC calculation (bottom left quadrant, northern hemisphere data). The
  difference between them is shown in the right half of each plot,
  which shows data from both hemispheres. The shock front is drawn as
  a contour at entropy $S=7\,k_B\,\mathrm{baryon}^{-1}$ and is colored
  for clarity. The results from other neutrino species behave nearly
  identically. These plots effectively replicate the 1D results in
  Figures~\ref{fig:1DFr} and \ref{fig:1DPrr}, but with a DO angular
  resolution between the 1D\_1x and 1D\_2x resolutions. MC results
  show a faster transition to forward-peaked distribution functions
  due to the limited angular resolution in the DO calculation.}
\label{fig:2DFr}
\end{figure*}
\begin{figure*}
  \includegraphics[width=\linewidth]{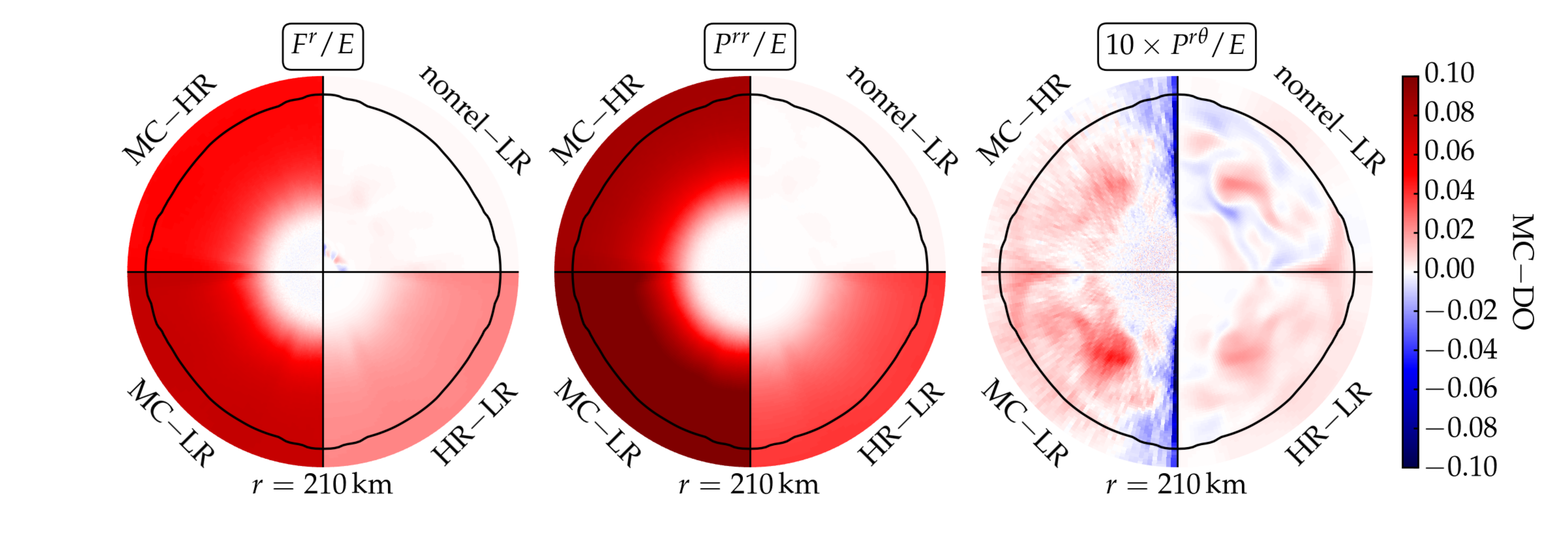}
  \caption{\textbf{2D Resolution and Relativity}. All plots show
    differences between angular moments of the energy-integrated
    electron neutrino radiation field using different calculations. In
    each plot, we show a comparison between the 2D\_HR\_native MC
    calculation and the 2D\_HR DO calculation (top left), between the
    2D\_HR\_native MC calculation and the 2D\_LR DO calculation
    (bottom left), between the 2D\_HR DO calculation and the 2D\_LR DO
    calculation (bottom right), and between the 2D\_LR\_nonrel DO
    calculation and the 2D\_LR DO calculation (top right). The left
    plot shows the radial component of the flux flux factor, the
    center plot shows the $rr$ component of the Eddington tensor, and
    the right plot shows the $r\theta$ component of the Eddington
    tensor. The shock front is drawn as a contour at entropy
    $S=7\,k_B\,\mathrm{baryon}^{-1}$. The left quadrants of each plot
    show that the DO error decreases with increasing angular
    resolution. The top right quadrant shows that special relativistic
    effects have a relatively small impact on the moments.}
  \label{fig:2Dmisc}
\end{figure*}
In Figure~\ref{fig:2DFr}, we plot the energy-integrated lab-frame
electron neutrino flux factor and Eddington tensor using the 2D\_HR DO
calculation and the 2D\_HR\_native MC calculation. These plots
effectively reproduce the 1D angular moment results in
Figures~\ref{fig:1DFr} and \ref{fig:1DPrr}. We again see that MC
results exhibit a quicker transition to a forward-peaked distribution,
that the errors in the second moment ($P^{rr}/E$) are larger than in
the first ($F^r/E$), and that the MC noise in the second moment is
smaller than in the first. The magnitude of the differences in
$P^{rr}/E$ of $\sim0.1$ at the shock can be compared to the 1D results
in Figure~\ref{fig:1Dresolution}. The 2D\_HR differences are between
the 1D\_1x and 1D\_2x differences, reflecting the fact that the 2D\_HR
angular resolution is between that of the 1D\_1x and 1D\_2x
calculations. We also demonstrate this resolution dependence in the
leftmost (for $F^r/E$) and center (for $P^{rr}/E$) plots of
Figure~\ref{fig:2Dmisc}. The top left quadrant of each shows the
difference between the moments calculated using the 2D\_HR DO
calculation and the 2D\_HR\_native MC calculation. The bottom
left quadrant shows the difference between the 2D\_LR DO calculation
and the same MC calculation. The differences are significantly smaller
for the higher resolution DO calculation, indicating that the DO
results are converging to the MC results with increasing resolution.

\begin{figure*}
\includegraphics[width=\linewidth]{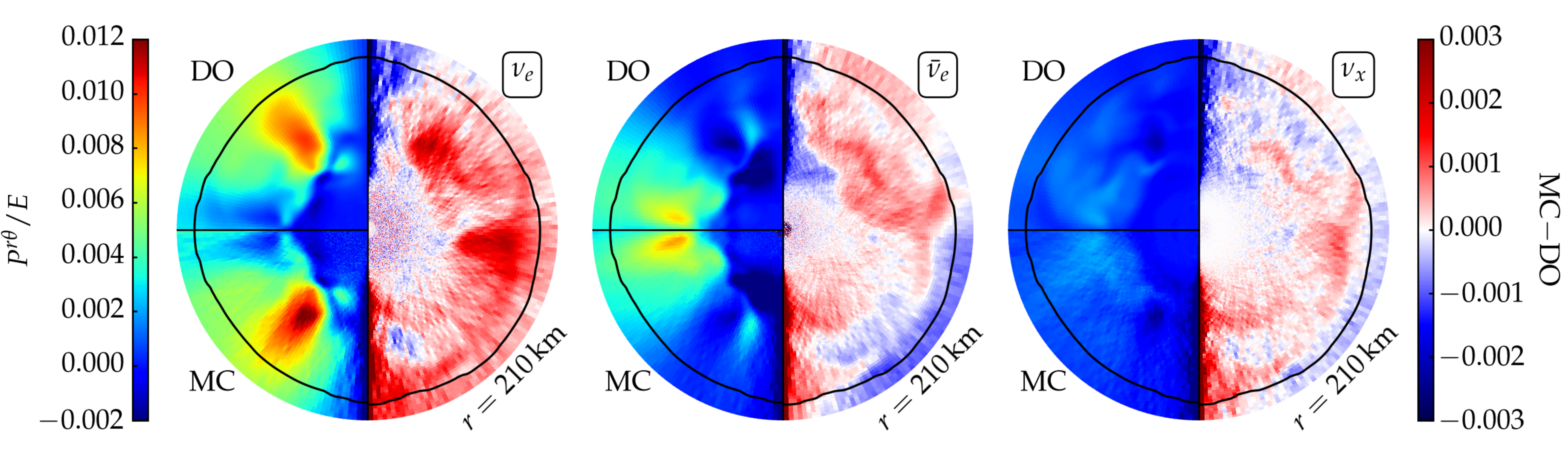}
\caption{\textbf{2D Eddington Tensor (Off-Diagonal).}  All plots show
  the energy-integrated lab-frame $r\theta$ component of the Eddington
  tensor. This is a sensitive probe of multi-dimensional anisotropy,
  as it is identically zero in 1D calculations. In each plot, we show
  the radiation moment computed using the 2D\_HR DO calculation (top
  left quadrant, northern hemisphere data) and the 2D\_HR\_native MC
  calculation (bottom left quadrant, northern hemisphere data). The
  left plot shows electron neutrinos, the center shows electron
  anti-neutrinos, and the right shows heavy lepton neutrinos. The
  difference between the MC and DO results is shown in the right half
  of each plot, which shows data from both hemispheres. The shock
  front is drawn as a contour at entropy
  $S=7\,k_B\,\mathrm{baryon}^{-1}$ and is colored for clarity. MC
  results show larger values of $P^{r\theta}$ due to limited angular
  resolution in the DO calculation (see right panel of
  Figure~\ref{fig:2Dmisc}).}
\label{fig:2DPrt}
\end{figure*}
Unlike in spherical symmetry, $P^{r\theta}$ is not identically zero in
axisymmetry and is thus a sensitive probe of multidimensional effects
on the radiation field. $P^{r\phi}$ and $P^{\theta\phi}$ are still
identically zero, since we do not consider azimuthal fluid
velocities. In Figure~\ref{fig:2DPrt}, we plot the energy-integrated
lab-frame $P^{r\theta}/E$ for all three neutrino species using the
2D\_HR DO calculation and the 2D\_HR\_native MC calculation. Since the
dominant neutrino propagation direction is radial, the off-diagonal
components of the pressure tensor are strongly correlated with the
corresponding flux. In this particular snapshot, $P^{r\theta}$ happens
to be overall mostly positive, and we find the morphology to be indeed
very similar to $F^\theta$ (not shown). Generally, both positive and
negative values are to be expected (see \citealt{nagakura:17a}). It is
interesting to note that the electron neutrino and electron
anti-neutrino plots have complementary hot spots. Within the
protoneutron star, non-radial neutrino fluxes are present due to
turbulent fluid carrying trapped neutrinos. Outside of the convective
zone of the PNS but still within the neutrinospheres, electron
neutrinos tend to diffuse away from regions of high electron chemical
potential while electron anti-neutrinos diffuse toward them. In tests
where the inner $105\,\mathrm{km}$ is excluded from the calculation,
the $P^{r\theta}$ distribution is much more uniform, suggesting that
the hot spots are due to a combination of anisotropic neutrinos from
the neutrinosphere interacting with multi-dimensional features in the
fluid background.

Once again, the MC and DO results for $P^{r\theta}$ look remarkably
similar. Unlike for the diagonal moments, much subtractive
cancellation occurs when computing $P^{r\theta}$, which in turn
requires a large number of MC particles to drive down the
noise. Similar to the other moments, the MC calculation tends to show
larger values of $P^{r\theta}$, since its effectively infinite angular
resolution is able to resolve finer angular structures. We demonstrate
this resolution dependence in the rightmost plot of
Figure~\ref{fig:2Dmisc}. The top left quadrant shows the difference
between the electron neutrino $P^{r\theta}$ from the 2D\_HR DO and
2D\_HR\_native MC calculations, while the bottom left quadrant
compares the 2D\_LR DO calculation to the same MC calculation. The
differences are significantly larger for the lower-resolution
calculation, indicating that the DO calculation is converging to the
MC result with increasing angular resolution.

\begin{figure*}
  \includegraphics[width=\linewidth]{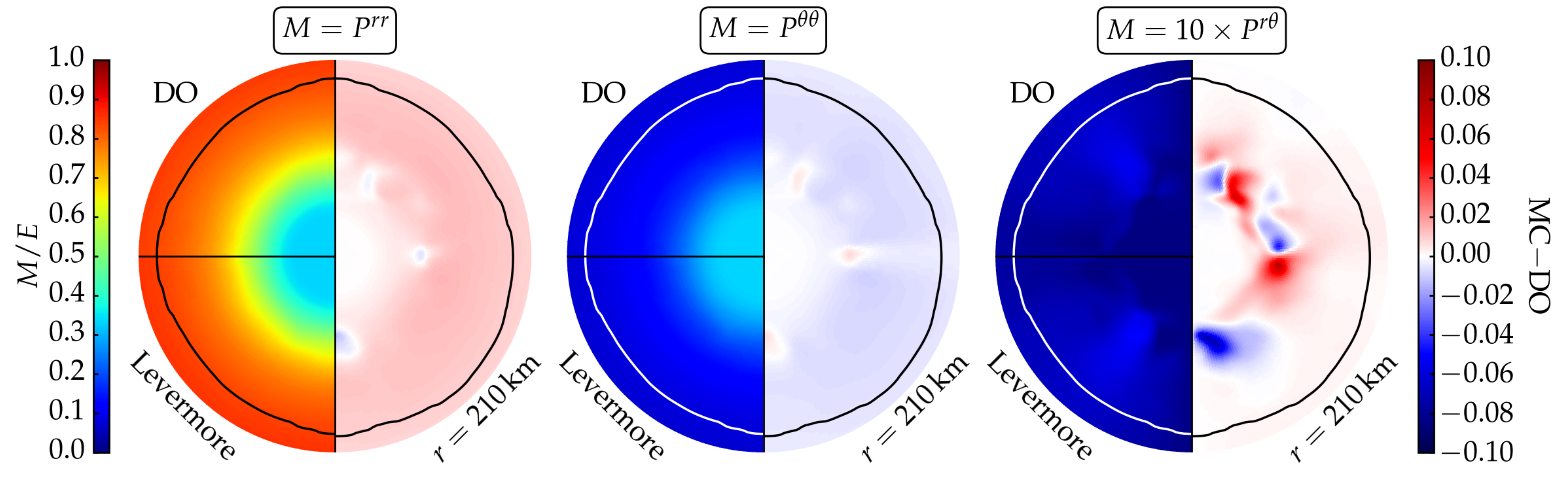}
  \caption{\textbf{2D Approximate Closures}. All plots show
    energy-integrated components of the Eddington tensor of the
    electron neutrino radiation field in the lab frame. The leftmost
    plot shows the $rr$ component, the center plot shows the
    $\theta\theta$ component, and the right plot shows the $r\theta$
    component, multiplied by 10 (including the differences) for
    clarity. In each plot, the top left quadrant shows results from
    the northern hemisphere of the 2D\_HR DO calculation. The bottom
    left shows results when the energy density and flux from the
    2D\_HR calculation are used to calculate the Eddington tensor
    component using the Levermore closure (also northern hemisphere
    data). The right half shows the difference between the two, and
    includes data from both hemispheres. The shock front is drawn as a
    contour at entropy $S=7\,k_B\,\mathrm{baryon}^{-1}$ and is colored
    for clarity. As in the 1D calculations, the Levermore closure is a
    good approximation for diagonal components, but it struggles for
    off-diagonal components. Other closures show slightly larger
    errors. Electron anti-neutrino and heavy lepton neutrino results
    behave similarly, except that the overall magnitude of
    $P^{r\theta}$ is smaller for heavy lepton neutrinos.}
  \label{fig:2Dclosure}
\end{figure*}
Figure~\ref{fig:2Dclosure} compares components of the electron
neutrino Eddington tensor computed by the 2D\_HR DO calculation to
those predicted by the Levermore closure using $E$ and $F^i$ from the
same DO calculation. We demonstrated in Section~\ref{sec:1Dcomparison}
that in our spherically symmetric snapshot, the Levermore closure
predicts $P^{rr}/E$ and $P^{\theta\theta}/E$ from only the flux factor
with an accuracy within 0.01 of the actual Eddington tensor
value. This result is reproduced in two dimensions for electron
neutrinos in Figure~\ref{fig:2Dclosure}. The leftmost and center plots
show $P^{rr}/E$ and $P^{\theta\theta}/E$, respectively. The top left
quadrant of each shows the moment computed directly from the 2D\_HR DO
calculation (same data as depicted in Figure~\ref{fig:2DFr}), and the
bottom left quadrant shows the same moment predicted by the Levermore
closure. They are visually identical, and the error plotted on the
right side of each plot shows a maximum error of $0.014$ in $P^{rr}/E$
and a maximum error of 0.0089 in $P^{\theta\theta}/E$. Though there is
some multi-dimensional structure in how accurately the Levermore
closure predicts the diagonal components, this effectively mirrors the
results of the 1D calculations. The rightmost plot shows
$P^{r\theta}/E$ (same data as in Figure~\ref{fig:2DPrt}), multiplied
by 10 for visibility on this color scale. The Levermore closure
predicts this component of the moment within 0.0077. This is large
compared to this component's maximum value of 0.012 and compared to a
difference of $\sim0.003$ between DO and MC results. Thus, though this
analytic closure has small relative errors for the diagonal components
of the Eddington tensor, it has difficulty accurately predicting the
small off-diagonal components in this CCSN snapshot. The Minerbo and
Janka closures show errors at smaller radii, but the extrema of these
errors are only slightly larger than those of the Levermore
closure. Other neutrino species behave very similarly, except that the
heavy lepton neutrino values for $P^{r\theta}$ (and hence errors) are
significantly smaller.

Ignoring special relativistic effects in radiation transport
calculations greatly simplifies the problem. Fluid velocities are also
generally only a few percent of the speed of light below the shock,
but are larger in 2D ($\sim0.037 c$) than in 1D ($\leq 0.015 c$). We
test the effects of ignoring fluid velocities in
Figure~\ref{fig:2Dmisc}, where we plot the difference between the
2D\_LR\_nonrel and 2D\_LR DO calculations. The error in $F^r/E$ and
$P^{rr}/E$ from ignoring velocities is much smaller than the
difference between MC and DO calculations or the difference between
resolutions. The only exception is in the convective region of the
protoneutron star, where the flux is determined entirely by the fluid
velocity because the neutrinos are trapped. The magnitude of this
error is at most comparable to the error induced by the coarse
resolution, and is significantly smaller than the error in all
components of the second moment predicted by the Levermore closure
(Figure~\ref{fig:2Dclosure}).

\begin{figure*}
  \centering
  \includegraphics[width=0.75\linewidth]{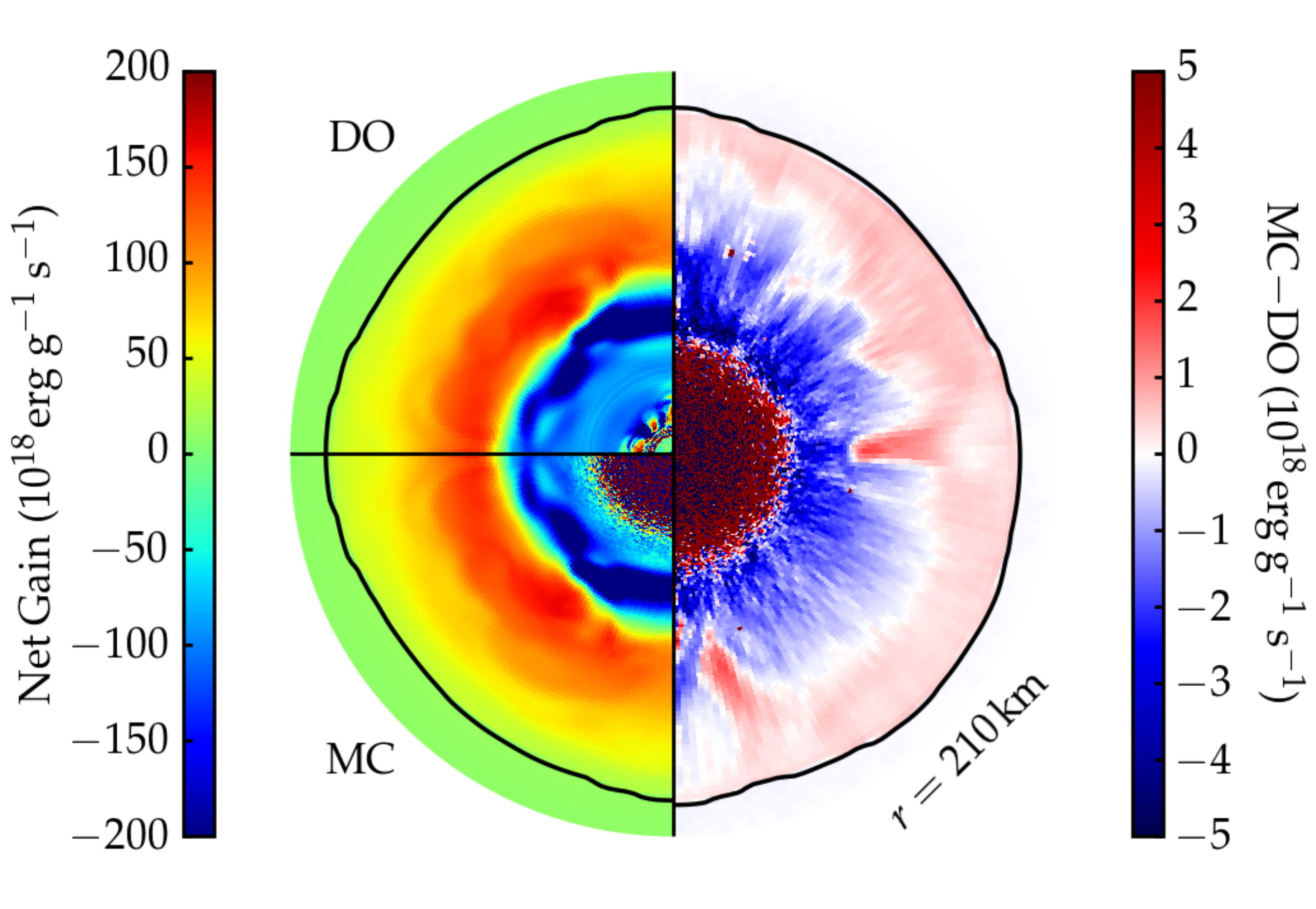}
  \caption{\textbf{2D Net Gain} Net gain ($\mathcal{H}-\mathcal{C}$)
    using the 2D\_HR DO calculation (top left, northern hemisphere
    data) and the 2D\_HR\_native MC calculation (bottom left, northern
    hemisphere data). The difference between the two is shown in the
    right half, and contains data from both hemispheres. The shock
    front is drawn as a contour at entropy
    $S=7\,k_B\,\mathrm{baryon}^{-1}$. The MC data shows slightly
    faster cooling in the cooling region, and slightly faster heating
    in the gain region. The MC data below $\sim70\,\mathrm{km}$ is
    dominated by noise.}
  \label{fig:2Dgain}
\end{figure*}
Finally, in Figure~\ref{fig:2Dgain}, we investigate how these
different transport schemes affect the actual heating and cooling
rates of the fluid. Once again, we show the results from the 2D\_HR DO
calculation in the top left quadrant, which outside the core appears
visually identical to the 2D\_HR\_native MC calculation results in the
bottom left quadrant. Just as in Figure~\ref{fig:1Dgain}, there is
significant statistical noise within the core, where neutrinos are
largely in equilibrium with the matter. We depict the difference
between these results on the right half of the plot. The MC results
show more rapid cooling in the cooling region and more rapid heating
in the outer regions of the gain layer, but only by a few percent of
the maximum net gain outside of the protoneutron star. This is similar
to the behavior of the lower-resolution 1D results in
Figure~\ref{fig:1Dgain}, where MC calculations predict a smaller gain
than do the DO calculations at $r\lesssim 125\,\mathrm{km}$ for the
1D\_2x calculation and at $r\lesssim 140\,\mathrm{km}$ for the 1D\_1x
calculations. The 2D MC calculation also predicts larger heating than
does the 2D\_HR DO calculation in regions of high inward velocity.
This is again an effect of the limited momentum-space resolution in
the DO calculations that make the two-grid approach somewhat diffusive
in angle and energy. This is to be expected given that the average
neutrino energies in these regions (Figure~\ref{fig:2DavgE}) are
higher in the MC calculations. Overall, excluding the noisy region in
the core, these errors are at most $\sim2\%$ of the amplitude of the
net gain curve in Figure~\ref{fig:1Dgain}.

The volume-integrated gain of the gain region (where there is net
heating under the shock) is $9.00\times10^{51}\,\mathrm{erg\,s}^{-1}$
in the 2D\_HR DO calculation and
$8.93\times10^{51}\,\mathrm{erg\,s}^{-1}$ in the 2D\_HR MC
calculation, which is only a $0.35\%$ difference. Compare to this the
relative error of the same quantity between the 2D\_LR and 2D\_HR DO
calculations of $1.3\%$, which is smaller than in the 1D resolution
comparison because our 2D resolutions are much more similar. Even so,
the errors from low resolution are significantly more significant than
the differences between the methods. The difference in integrated
heating rate between the 2D\_LR and 2D\_LR\_nonrel DO calculations is
$2.0\%$. This is larger than in the 1D relativity comparison because
fluid velocities under the shock are larger in the 2D calculations
than in the 1D calculations due to convective motion. Note that the
integrated heating rate should not be compared with the 1D results
because the fluid profiles are significantly different.

%%%%%%%%%%%%%%%
% CONCLUSIONS %
%%%%%%%%%%%%%%%
\section{Conclusions}
\label{sec:conclusions}

Neutrinos dominate energy transport in CCSNe and are a vital component
of the CCSN explosion mechanism. It is therefore imperative to ensure
neutrinos are simulated accurately in CCSN models. One means of
checking the accuracy of a computational method is comparing against
another accurate method. The grid-based discrete ordinates (DO) method
and particle-based Monte Carlo (MC) method both solve the full
transport problem in very different ways. We perform the first
detailed multi-dimensional comparison of special relativistic
Boltzmann-level neutrino transport codes in the context of
core-collapse supernovae using the grid-based discrete ordinates (DO)
code of \cite{nagakura:17a} (NSY) and the particle-based Monte Carlo
(MC) {\tt Sedonu} code. We verify that both methods converge to the
same result in the limit of large MC particle count and fine DO
momentum-space resolution under the assumption of a static fluid
background in spherical symmetry and in axisymmetry. This provides
confidence in the accuracy of the results from these two completely
different approaches.

We demonstrate an agreement of the average neutrino energy to within
$\sim 0.1$ MeV for 1D calculations and $\sim0.5$ MeV for coarser 2D
calculations everywhere in the simulation domain for all three
simulated neutrino species. We also demonstrate exquisite agreement in
the local spectra of all three species. We find that numerical
diffusion from a coarse momentum-space resolution dominates these
small errors, though smaller errors result from finite spatial
resolution and from the Monte Carlo random walk approximation.

MC transport computes angular moments of the distribution function
with great accuracy when the moments are computed natively during the
calculation as opposed to post-processed from an angular grid. The DO
method requires a very high angular resolution of about 40 points in
the polar direction to compute these moments with similar accuracy in
1D calculations, which is currently not possible in 2D calculations
and certainly not possible in 2D time-dependent simulations. The MC
method, however, requires a large number of particles to be simulated
to reach low noise levels in moments that exhibit subtractive
cancellation (i.e., $F^i$ in optically-deep regions, $P^{r\theta}$ in
2D calculations). The present 2D calculation simulated 63 billion
particles and still show some noise in these quantities.

The approximate two moment radiation transport scheme is significantly
more efficient than either DO or MC transport by evolving on the the
energy density and flux. However, this method requires an ad-hoc
closure relation to complete the system of equations by making
estimates of higher-order moments. We evaluate the performance of the
Levermore, Janka, and Minerbo closures in predicting the second
angular moments from the first. In the 1D calculations, the Levermore
closure performs best, with an error comparable to the differences
between the highest-resolution DO results and the MC results. In 2D
calculations, this closure performs comparably well when predicting
diagonal components of the second moment, but this accuracy is not
sufficient for determining the very small off-diagonal
components. Though careful tests would be required to assess the
importance of these small off-diagonal components, these components
reflect the multi-dimensional nature of CCSN dynamics.

Finally, we find that the difference in local heating and cooling
rates between the DO and MC methods is at most $2\%$ of the amplitude
of the net gain curve in the cooling region of the CCSN in both 1D and
2D calculations. The volume-integrated gain in the gain region (where
there is net heating under the shock), however, differs by only
$0.3\%$ in the highest resolution 1D calculations and by $0.4\%$ in
the highest resolution 2D calculations. In these cases, the DO and MC
schemes share the same energy resolution, but the MC scheme has
effectively infinite angular resolution. The differences in the same
quantities due to changing the DO energy (and angular) resolution are
larger than $1\%$ in both 1D and 2D calculations, indicating that
neutrino energy resolution is the dominant source of real error. Since
both the MC and DO methods rely on opacities and emissivities
discretized into energy bins, both suffer from this error. The errors
in radiation quantities (energy density, angular moments, and average
neutrino energies) below the shock is dominated by the finite
momentum-space resolution of the DO calculations and statistical noise
and limited energy resolution of the MC calculations.

Though it is important to simulate all physics relevant to the CCSN
mechanism, the numerical resolution can pose a significantly larger
threat to simulation accuracy than a lack of physical elements. We
test the effects of ignoring special relativistic Lorentz
transformation of neutrinos and find it to be severely sub-dominant to
errors induced by low momentum-space resolution at the resolutions we
use. The diagonal components of the Eddington tensor in the
low-resolution DO calculations show resolution-induced errors of
$\sim15\%$, and even the highest-resolution 1D calculations (which
would be impossible in 2D) show errors of $\sim3\%$. This underscores
the need for resolution tests in interpreting results of simulation
results.

Though this study inspires much more confidence in both methods, we
must mention several caveats. The largest is that
opacities and emissivities are an extremely important component of
radiation transport. In order to facilitate a detailed comparison of
the methods themselves, we carefully configure both codes to use
identical opacities at each spatial location and neutrino energy
bin. However, we do not compare the effects of different
approximations and physical processes present in the opacities that
may overwhelm the small differences in the results between these
codes.

Second, we must emphasize that our calculations are performed under
the assumption of an unchanging fluid background and flat metric at
one particular stage in the CCSN evolution. At different stages,
especially during early postbounce prompt convection and the shock
revival phase, the matter distribution and hence neutrino radiation
fields are significantly different and would benefit from a similar
analysis. We also use an approximate treatment of pair processes and
neglect electron scattering. These simplifications are made to bring
both codes to the same level, where we could be sure that they are
simulating the same physics with the same level of approximation. This
allows an isolated evaluation of the relative performance of both
transport methods, but neglects many components of physics that should
be included in realistic dynamical CCSN simulations.

The impact of the time-dependent features of the radiation field on
the fluid dynamics will be the next necessary step in verifying
neutrino radiation hydrodynamics codes. A similar careful verification
of the choice and implementation of different neutrino interactions,
the resolution and discretization scheme (including mesh geometry and
refinement), the treatment of gravity, and the numerical hydrodynamics
scheme would also greatly benefit the interpretation of simulation
results. We leave this broader task of evaluating multi-dimensional
time-dependent radiation hydrodynamics simulations of CCSNe to future
work.

We release the data input to and output by both codes at
\url{http://www.stellarcollapse.org/MCvsDO}. The {\tt Sedonu} code is
also open source and available at
\url{https://bitbucket.org/srichers/sedonu.git}, along with a set of
ready-to-run input data and parameter files to run the calculations in
this paper. This {\tt Sedonu} release contains many performance,
usability, and flexibility changes implemented since previous
releases. In addition, we incorporate a special relativistic,
time-independent version of the MC random walk approximation that
enables {\tt Sedonu} to efficiently calculate neutrino transport
through regions of arbitrarily large optical depth.

%%%%%%%%%%%%%%%%%
% MISCELLANEOUS %
%%%%%%%%%%%%%%%%%
\acknowledgments We would like to acknowledge Ryan Wollaeger, Kendra
Keady, Adam Burrows, David Radice, Luke Roberts, and Yuki Nishino for
many insightful discussions of radiation transport methods. S.~R. was
supported by the National Science Foundation (NSF) Blue Waters
Graduate Fellowship. This research is part of the Blue Waters
sustained-petascale computing project, which is supported by the
National Science Foundation (Grants No. OCI-0725070 and
No.\ ACI-1238993) and the State of Illinois. Blue Waters is a joint
effort of the University of Illinois at Urbana-Champaign and its
National Center for Supercomputing Applications.  This work benefited
from access to the NSF XSEDE network under allocation TG-PHY100033 and
to Blue Waters under PRAC award no.\ ACI-1440083. H.~N. is supported
in part by JSPS Postdoctoral Fellowships for Research Abroad
No.~27-348. C.~D.~O. is supported in part by the International
Research Unit of Advanced Future Studies at the Yukawa Institute for
Theoretical Physics. This work is furthermore partially supported by
the Sherman Fairchild Foundation and NSF under award nos.\ TCAN
AST-1333520, CAREER PHY-1151197, and PHY-1404569. This work is
supported by a Grant-in-Aid for Scientific Research from the Ministry
of Education, Culture, Sports, Science, and Technology (MEXT) of Japan
(15K05093, 16H03986, 26104006) and the HPCI Strategic Program of
Japanese MEXT and K computer at the RIKEN and Post-K project (Project
ID: hpci 170304, 170230, 170031). J.~D.~acknowledges support from the
Laboratory Directed Research and Development program at Los Alamos
National Laboratory (LANL).  Work at LANL by S.~R.\ and J.~D.\ was
done under the auspices of the National Nuclear Security
Administration of the US Department of Energy. This paper has been
assigned Yukawa Institute for Theoretical Physics report number
YITP-17-61.

\bibliography{../bibliography/sn_theory_references.bib,../bibliography/methods_references.bib,../bibliography/radiation_transport_references.bib,../bibliography/NSNS_NSBH_references.bib,../bibliography/bh_formation_references.bib,../bibliography/stellarevolution_references.bib,../bibliography/nu_oscillations_references.bib,../bibliography/eos_references.bib,../bibliography/grb_references.bib,../bibliography/nucleosynthesis_references.bib,../bibliography/cs_hpc_references.bib,../bibliography/nu_interactions_references.bib}

%%%%%%%%%%%%%%%%%%%%%%
% BOLTZMANN APPENDIX %
%%%%%%%%%%%%%%%%%%%%%%
\begin{appendix}
\section{General Relativistic Boltzmann Equation}
\label{app:boltzmann}

The NSY code solves the conservative form of the general relativistic
Boltzmann equation, which can be written as \citep{nagakura:17a}
\begin{equation}
\begin{aligned}
 &  \frac{1}{\sqrt{-g}} \left. \frac{\partial}{\partial x^{\alpha}} \right|_{\mathbf{q}} 
  \left[\sqrt{-g} f \left(e^\alpha_{(0)} + \sum^{3}_{i=1} \ell_{(i)} e^{\alpha}_{(i)}\right) \right]\\
  & - \frac{1}{\epsilon^2} \frac{\partial}{\partial \epsilon}\left(\epsilon^3 f \omega_{(0)} \right)
  + \frac{1}{\sin\bar{\theta}} \frac{\partial}{\partial \bar{\theta}} \left(\sin\bar{\theta} f \omega_{(\bar{\theta})} \right)
  + \frac{1}{\sin^2 \bar{\theta}} \frac{\partial}{\partial \bar{\phi}} \left(f \omega_{(\bar{\phi})}\right)\,\, = S_\mathrm{rad}  ,
\label{eq:basicBoltz}
\end{aligned}
\end{equation}
where $S_{\rm{rad}}$ describes the collision term for neutrino-matter
interactions, $g$ is the determinant of the 4-dimensional metric, and
$x^{\alpha}$ are the spacetime coordinates. $e^{\alpha}_{(\mu)}$ ($\mu
= 0, 1, 2, 3$) denote a set of tetrad bases for a local orthonormal
frame on 4-dimensional manifolds. In the present study, we assume that
the spacetime is flat, so we simply use spherical-polar coordinates,
i.e., $x^0=t$, $x^1=r$, $x^2=\theta$, $x^3=\phi$, $e_{(0)}=\hat{t}$,
$e_{(1)}=\hat{r}$, $e_{(2)}=\hat\theta$, $e_{(3)}=\hat\phi$, and
$g=-r^2\sin\theta$. $x^1$,$x^2$, and $x^3$ contain the same
information as $\mathbf{x}$ in Section~\ref{sec:transmethods}. The
neutrino momentum space is also written in spherical-polar coordinates
$q^i$ ($i=1,2,3$). $q^1=\epsilon\equiv -p_\alpha e_{(0)}^\alpha$ is
the energy of a neutrino with four-momentum $p^\alpha$,
$q^2=\bar\theta$ is the polar direction angle with respect to
$\hat{r}$, and $q^3=\bar{\phi}$ is the azimuthal angle with respect to
$\hat\theta$. $q^2$ and $q^3$ contain the same information as
$\mathbf{\Omega}$ in Section~\ref{sec:transmethods}. The derivative in
the first term of Equation~\ref{eq:basicBoltz} is evaluated while
holding the momentum coordinates constant. The direction cosines
$\ell_{i}$ are equivalent to $\mathbf{\Omega}\cdot \mathbf{e}_{(i)}$
in Section~\ref{sec:transmethods} evaluated in the lab frame and can
be written as
\begin{equation}
\begin{aligned}
  \ell_{(1)} &= {\rm cos} \hspace{0.5mm} \bar{\theta}\,\,,\\
  \ell_{(2)} &= {\rm sin} \hspace{0.5mm} \bar{\theta}   {\rm cos} \hspace{0.5mm} \bar{\phi}\,\,,\\
  \ell_{(3)} &= {\rm sin} \hspace{0.5mm} \bar{\theta}   {\rm sin} \hspace{0.5mm} \bar{\phi}\,\,.
  \label{eq:el}
\end{aligned}
\end{equation}
The geometric coefficients $\omega_{(0)}$, $\omega_{(\bar{\theta})}$,
$\omega_{(\bar{\phi})}$ given in \cite{nagakura:17a} reduce in flat
spacetime to
\begin{equation}
\begin{aligned}
  \omega_{(0)} &= 0\,\,, \\
  \omega_{(\bar{\theta})} &= - \frac{{\rm sin} \bar{\theta} }{r}\,\,,\\
  \omega_{(\bar{\phi})} &= - \frac{ {\rm cot}\theta }{r} {\rm sin}^3 \bar{\theta} \hspace{0.5mm} {\rm sin} \bar{\phi}\,\,.
  \label{eq:Omega_flat}
\end{aligned}
\end{equation}

The NSY code uses Lagrangian remapping grids (LRG) and laboratory
fixed grids (LFG) in the fluid rest frame and the lab frame,
respectively, to discretize the neutrino momentum space
\citep{nagakura:14}. The LFG are constructed so as to have an
isotropic energy grid in the lab frame, while the LRG are constructed
so as to have an isotropic energy grid in the fluid rest frame and a
propagation angle-dependent energy grid in the lab frame. Here,
isotropic means that each angular bin sees the same energy grid. This
two-grid technique is essential to treating special-relativistic
effects in full generality in the DO method.

\section{Monte Carlo Random Walk Approximation}
\label{app:montecarlo}

In regions where the scattering optical depth $\tau_s = \kappa_s l$ is
large, where $l$ is the relevant length scale, direct MC
radiation transport becomes very inefficient. The path length between
scattering events is very small, so a great deal of computer time is
spent performing these scattering events while there is little actual
movement of energy and lepton number. In these regions, the neutrino
transport is very well approximated as a diffusion process, a fact
which we use to accelerate the computation.

In the past, MC neutrino transport schemes have excluded the
inner regions of high optical depth in favor of an inner boundary
condition \citep{janka:91} or have employed the discrete diffusion
MC approximation in these regions
\citep{densmore:07,abdikamalov:12}. In order to keep the neutrino
motion free of any specific grid geometry and to prevent a hard
spatial boundary between two algorithms, we instead choose to
implement the MC random walk approximation
\citep{fleck:84}. This treats neutrino motion over a specified
distance as a diffusive process, and relies on the assumption of
isotropic, elastic scattering. In our implementation, we also assume
the fluid is unchanging in space and in time during each diffusion
step. Here we modify the method of \cite{fleck:84} to treat static
fluid backgrounds with relativistic fluid velocities.

The approximation accelerates MC transport in regions of high
scattering optical depth using a solution to the diffusion equation:
\begin{equation}
  \partial_t \psi(\mathbf{r},t) = D \nabla^2 \psi(\mathbf{r},t)\,\,.
\end{equation}
The diffusion constant can be shown to be $D=c/3\kappa_s$ by comparing
the solution to the diffusion equation on an infinite uniform
background given initial conditions $\psi(\mathbf{r},0) =
\delta^3(\mathbf{r})$ to the solution of a random walk process
starting at $\mathbf{r}=0$ with step sizes determined by the
scattering opacity $\kappa_s$ \citep{fleck:84,chandrasekhar:43}. In
the context of MC radiation transport, the solution
$\psi(\mathbf{r},t)$ represents the probability density of the
neutrino being at location $\mathbf{r}$ at time $t$.

\begin{figure}
  \center
  \includegraphics[width=0.5\linewidth]{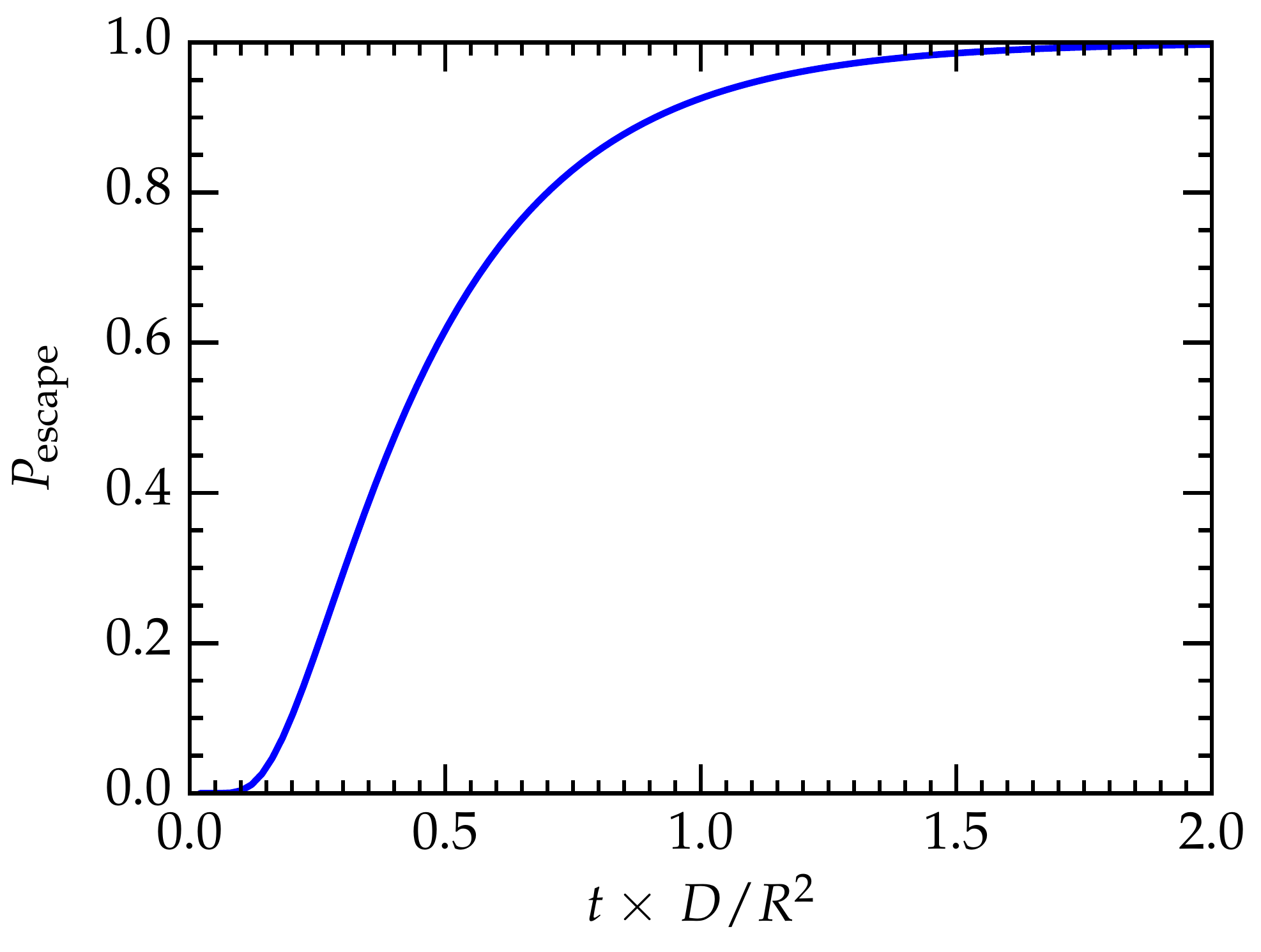}
  \caption{Probability of escape from a sphere of radius R and
    diffusion constant D after time t
    (Equation~\ref{eq:Pescape}). Inverse transform sampling is applied
    to this function to randomly sample the time it takes a neutrino
    to reach the edge of a diffusion sphere in the MC random
    walk approximation.}
  \label{fig:Pescape}
\end{figure}

Using the diffusion equation with this diffusion constant, we now
specify a sphere of radius $R$ in the comoving frame and derive the
probability that a neutrino has escaped from the sphere after a
certain time $t$. To do this, we again solve the diffusion equation,
but this time with the boundary condition $\psi(R,t)=0$ to indicate
that we are interested only in the \textit{first} time a neutrino
leaves the sphere and we do not allow neutrinos to leave and then
re-enter the sphere. This can be solved via separation of variables
and Sturm-Liouville orthogonality conditions to arrive at
\begin{equation}
  \psi(r,t>0) = \sum_{n=1}^\infty \frac{n}{2R^2} \frac{\sin(n\pi r/R)}{r} \exp\left[-\left(\frac{n\pi}{R}\right)^2 D t\right]\,\,.
\end{equation}
 The probability
that a neutrino has escaped the sphere after time t is
represented by the volume integral of the diffusion solution
(Figure~\ref{fig:Pescape}):
\begin{equation}
  \label{eq:Pescape}
  \begin{aligned}
    P_\mathrm{escape}(R,t>0) &= 1-\int_0^R \psi(r,t) 4\pi r^2 dr \\ &=
    1-2\sum_{n=1}^\infty (-1)^{n-1}
    \exp\left[-\left(\frac{n\pi}{R}\right)^2Dt\right]\,\,.\\
    \end{aligned}
\end{equation}
This solution is plotted in Figure~\ref{fig:Pescape}.

The diffusion equation is acausal in that there is a finite
probability of a neutrino escaping at times less than the light travel
time to the edge of the sphere. Because of this, we set
$P_\mathrm{escape}(R,t<R/c)=0$. We can also use the escape probability
at $t=R/c$ as an estimate of the accuracy of the approximation. We
only use the random walk approximation when
\begin{equation}
  \label{eq:RWtolerance}
  P_\mathrm{escape}(R,R/c) < tol\,\,.
\end{equation}
In this study, we use $tol=10^{-3}$, which corresponds to only using
the random walk approximation when the scattering optical depth of the
sphere is $\kappa_s R \geq 12$.

We tabulate $P_\mathrm{escape}(R,t)$, which can then be inverted via
inverse transform sampling (e.g., \citealt{haghighat:15}) to randomly
sample the escape time $t_\mathrm{esc}$. The table extends over the
range of $0 \leq \chi \leq \chi_\mathrm{max}$ using 100 evenly spaced
points in $\chi$ where $\chi = Dt/R^2$ and, in our calculations,
$\chi_\mathrm{max}=2$ (corresponding to $P_\mathrm{escape}=0.997$). We
evaluate the first 1000 terms in the sum in Equation~\ref{eq:Pescape},
which is far more than is necessary for a converged solution, but
tabulating $P_\mathrm{escape}$ is a very cheap one-time calculation.

We restrict the lab-frame radius of the sphere $R_\mathrm{lab}$ to the
largest length scale between (a) the distance to the nearest grid cell
boundary and (b) 1\% of the grid cell's smallest dimension. However,
since the sphere is defined in the fluid rest frame, its size must be
further limited when the fluid is moving, since the sphere is
effectively advected. The largest restriction occurs in the event that
the displacement of the neutrino from its starting position to the
surface of the sphere is parallel to the fluid velocity, so we will
use this worst-case scenario to set the sphere size limiter. The
four-vector $d_\mathrm{com}=\{t,\mathbf{R}\}$ connecting the
neutrino's initial and final positions in the lab frame can be
Lorentz-transformed to give the displacement vector in the lab frame
$d_\mathrm{lab}=\{\gamma(t+vR/c^2),\gamma(\mathbf{R}+\mathbf{v}t)\}$. The
longest diffusion time the numerical scheme will allow is
$t_\mathrm{max}=R^2\chi_\mathrm{max}/D$, resulting in a maximum
lab-frame displacement of $R_\mathrm{lab,max}=\gamma R(1+R v
\chi_\mathrm{max}/D)$. Inverting this, we set the comoving-frame
radius to
\begin{equation}
  R = \frac{2R_\mathrm{lab}}{\gamma}\left(1+\sqrt{1+\frac{4R_\mathrm{lab}v\chi_\mathrm{max}}{\gamma D}}\right)^{-1}\,\,.
  \label{eq:Rlimited}
\end{equation}

The comoving frame neutrino energy remains the same throughout the
process, since the scattering is assumed to be elastic. Absorption
happens continuously throughout the diffusion process. The packet
energy is decreased according to
\begin{equation}
  E_p (t) = E_p(0) \exp[-\kappa_a c t]\,\,,
  \label{eq:expdecay}
\end{equation}
and $E_p(0)-E_p(t)$ is added to the fluid energy to account for
neutrino absorption. The comoving frame packet energy averaged over
the diffusion time is
\begin{equation}
  \begin{aligned}
    \bar{E}_p &= \frac{1}{t} \int_0^t E_{p,0} e^{-\kappa_a c t'} dt' \\
    &= \frac{E_{p,0}}{ct \kappa_a}\left(1-e^{-\kappa_a c t}\right)\,\,.
  \end{aligned}
\end{equation}
If neutrino packets are created assuming the fluid emits for a time of
$\delta t_\mathrm{emit}$, this means that the neutrino contributes
$\bar{E}_p t / \delta t_\mathrm{emit}$ to the fluid cell's
steady-state radiation energy content. Averaged over the diffusion
process, most of the neutrino energy is distributed isotropically in
direction. However, there is a small asymmetry due to the fact that
the neutrino ends up at one point on the edge of the diffusion
sphere. Averaged over the duration of the diffusion process, for a
neutrino packet with energy $E_p$, there is a net energy flux of $E_p
R/ct$ in the direction of the final displacement vector while $E_p (ct
- R)/ct$ is distributed isotropically in direction.

With the theoretical groundwork complete, we now describe the random
walk algorithm itself. A comoving frame diffusion sphere size $R$ is
first chosen according to Equation~\ref{eq:Rlimited}. If the
scattering optical depth $\kappa_s R$ is sufficiently large
(Equation~\ref{eq:RWtolerance}), the time the neutrino takes to reach
the edge of the sphere $t$ is sampled from
Equation~\ref{eq:Pescape}. A location at the edge of the
comoving-frame sphere is randomly uniformly chosen, the displacement
4-vector $\{t,\mathbf{R}\}$ is Lorentz transformed into the lab frame,
and the neutrino is moved this distance. The new comoving neutrino
direction is chosen uniformly in the $2\pi$ steradians moving strictly
away from the diffusion sphere. The neutrino packet energy is
decreased due to absorption according to Equation~\ref{eq:expdecay}
and the absorbed energy is counted toward fluid heating. Comoving
radiation energy in the amount of $\bar{E}_p R / c\delta
t_\mathrm{emit}$ moving in the direction of the final displacement is
Lorentz transformed into the lab frame and accumulated into the
distribution function. The remaining $\bar{E}_p (ct-R)/c\delta
t_\mathrm{emit}$ of comoving radiation energy is divided evenly into
$N$ pieces, each is assigned an isotropically uniform random direction
in the comoving frame, is Lorentz transformed into the lab frame, and
is accumulated into the distribution function. This allows us to
self-consistently treat both the isotropic and directional components
of the radiation field without making reference to a particular grid
structure. In this work, we found that $N=10$ is a reasonable
compromise between code performance and noise in the resulting
radiation field.

\section{Comparison Details: Angular Moment Calculations}
\label{app:comparisonmethods}

The DO and MC methods are very different, so care is required to make
meaningful comparisons between the two codes. The NSY code evolves the
distribution function $f$, while \texttt{Sedonu} calculates the amount of
neutrino energy in each spatial-energy-direction cell in non-native
calculations. The \texttt{Sedonu} distribution function value at the bin center
$(r_a,\theta_b,\epsilon_k,\bar\theta_m,\bar\phi_n)$ is calculated
using
\begin{equation}
  f_{\mathrm{Sedonu},abkmn} = \frac{\varepsilon_{akmn}}{V_{ab} \Delta(\frac{\epsilon^3}{3})_k \Delta \cos(\bar\theta)_m \Delta\bar\phi_n}\,\,,
\end{equation}
where $\varepsilon_{akmn}$ is the total neutrino energy content (units
of ergs) in spatial-direction-energy bin $\{akmn\}$, $V_a$ is the
spatial volume of the grid cell in the lab frame, $\bar\theta$ and
$\bar\phi$ are the neutrino direction angles in the lab frame, and
$\epsilon$ is the neutrino energy in the comoving frame.

In the 1D simulations where \texttt{Sedonu} collects radiation information on
angular bins rather than native moments, we take care to ensure that
the post-processing for the two codes are equivalent. For both \texttt{Sedonu}
and the NSY code, the distribution function is linearly interpolated
to $\widetilde{f}$ on identical fine angular grids in
$\{\cos(\bar{\theta}),\bar{\phi}\}$ of $\{80,40\}$ zones,
respectively. Throughout this section, the subscript ($a$) refers to
the spatial mesh index, the subscript ($k$) refers to the neutrino
energy bin index, the subscripts ($mn$) refer to the direction indices
on the coarse direction grid used in the transport calculation, the
subscripts ($pq$) refer to the direction indices on the
high-resolution post-processing angular grid, and the superscripts
($ij$) refer to directions (i.e., $r$, $\theta$, or $\phi$) in the lab
frame.

The specific energy density (lab frame energy density per
comoving-frame neutrino energy) is computed as a sum over the coarse
grid for \texttt{Sedonu} and over the fine grid in the NSY code. This takes
advantage of the fact that \texttt{Sedonu} computes energy content directly and
does not introduce interpolation error into the \texttt{Sedonu} results:
\begin{equation}
  \begin{aligned}
    E_{\epsilon,ak,\mathrm{Sedonu}} &= \frac{1}{V_a\Delta\left(\frac{\epsilon^3}{3}\right)_k}\sum_m \sum_n \varepsilon_{akmn}\,\,,\\
    E_{\epsilon,ak,\mathrm{DO}} &= \epsilon_k  \sum_p \sum_q \widetilde{f}_{akpq} \Delta \cos(\bar{\theta}_{\nu})_p \Delta (\bar{\phi}_\nu)_q\,\,.
  \end{aligned}
\end{equation}
For both \texttt{Sedonu} and the NSY code, the higher-order moments are evaluated as
\begin{equation}
  \begin{aligned}
  F_{\epsilon,ak}^{i} &= \epsilon_k \sum_p \sum_q \widetilde{f}_{akpq} (\mathbf{\Omega}_{pq}\cdot \mathbf{e}_{(i)}) \Delta \cos(\bar{\theta})_p \Delta (\bar{\phi})_q\,\,,\\
  P_{\epsilon,ak}^{ij} &= \epsilon_k \sum_p \sum_q \widetilde{f}_{akpq} (\mathbf{\Omega}_{pq}\cdot \mathbf{e}_{(i)}) (\mathbf{\Omega}_{pq}\cdot \mathbf{e}_{(j)}) \Delta \cos(\bar{\theta})_p \Delta (\bar{\phi})_q\,\,.
  \end{aligned}
\end{equation}
The energy-integrated moments $M=\{E,F^i,P^{ij}\}$ are computed using
\begin{equation}
  M_a = \sum_k \Delta\left(\frac{\epsilon^3}{3}\right)_k M_{\epsilon,ak}\,\,.
\end{equation}
Finally, the average neutrino energy is computed using
\begin{equation}
  \bar{\epsilon}_a = \frac{E_a}{N_a}\,\,.
\end{equation}

\section{Comparison Details: Neutrino Reaction Rates}
\label{app:sourceterms}

The three source terms on the right hand side of
Equation~\ref{eq:boltzmann} each encapsulate multiple processes, and
are grouped into the mathematical nature of each term. In both \texttt{Sedonu}
and the NSY code, all of these source terms are evaluated in the
comoving frame. Details of how the NSY code computes reaction rates
are explained by \cite{bruenn:85} and \cite{sumiyoshi:05}.

The emission and absorption term takes the form of
\begin{equation}
  \left[\frac{\partial f}{c \partial t}\right]_\mathrm{em-abs} = R_\mathrm{em}(\epsilon)(1-f) - R_\mathrm{abs}(\epsilon)f\,\,,
\end{equation}
where $R_\mathrm{em}$ and $R_\mathrm{abs}$ are the
emission and absorption reaction rates, respectively. \texttt{Sedonu} takes
advantage of the concept of stimulated absorption to account for
final-state neutrino blocking \citep{brt:06}, in which the effective
absorption reaction rate is $\widetilde{R}_\mathrm{abs} =
R_\mathrm{abs} + R_\mathrm{emis}$. This removes the need to treat
final-state blocking explicitly in the neutrino emission process.

The scattering term accounts for neutrinos scattering into and out of
a given direction according to
\begin{equation}
  \left[\frac{\partial f}{c \partial t}\right]_\mathrm{scat} = \int
  d\mathbf{\Omega}' \int d\left(\frac{\epsilon'^3}{3}\right) \left[
    R_\mathrm{scat}(\epsilon',\epsilon,\mathbf{\Omega}'\cdot\mathbf{\Omega})f'(1-f)
    -
    R_\mathrm{scat}(\epsilon,\epsilon',\mathbf{\Omega}\cdot\mathbf{\Omega}')f(1-f')\right]\,\,.
\end{equation}
The primed variables are the neutrino final-state quantities and
$R_\mathrm{scat}$ is the scattering reaction rate. Both \texttt{Sedonu} and the
NSY code assume isoenergetic scattering, so the scattering reaction
rate becomes $R_\mathrm{scat}(\epsilon', \epsilon,
\mathbf{\Omega}'\cdot\mathbf{\Omega}) = \delta(\epsilon, \epsilon')
\widetilde{R}_\mathrm{scat}(\epsilon,\mathbf{\Omega}\cdot\mathbf{\Omega}')$.
Under this assumption, the scattering source term reduces to
\begin{equation}
  \left[\frac{\partial f}{c \partial t}\right]_\mathrm{scat} = \int
  d\mathbf{\Omega}' \widetilde{R}_\mathrm{scat}(\epsilon,\mathbf{\Omega}\cdot\mathbf{\Omega}') (f' - f)\,\,.
\end{equation}
\texttt{Sedonu} uses $\int d\mu \widetilde{R}_\mathrm{scat}(\epsilon,\mu)$,
where $\mu=\mathbf{\Omega}\cdot\mathbf{\Omega}'$, as the scattering
opacity directly.

Finally, pair annihilation and neutrino bremsstrahlung source terms
take the form of
\begin{equation}
  \left[\frac{\partial f}{c \partial t}\right]_\mathrm{pair-brem} = \int
  d\bar{\mathbf{\Omega}} \int d\left(\frac{\bar{\epsilon}^3}{3}\right) \left[
    R_\mathrm{pair-brem,emis}(\epsilon,\bar{\epsilon},\mathbf{\Omega}\cdot\bar{\mathbf{\Omega}})(1-f)(1-\bar{f})
    -
    R_\mathrm{pair-brem,abs}(\epsilon,\bar{\epsilon},\mathbf{\Omega}\cdot\bar{\mathbf{\Omega}})f\bar{f}\right]\,\,.
\end{equation}
The barred variables are the neutrino anti-species quantities and
$R_\mathrm{pair-brem,emis}$ is the reaction rate for pair and
bremsstrahlung processes. In order to ensure the same assumptions go
into both radiation transport schemes, the NSY code calculates these
reactions assuming the anti-species is isotropic, i.e.,
\begin{equation}
  \begin{aligned}
  \bar{f}_\mathrm{iso} &= \frac{1}{4\pi}\int d\bar{\mathbf{\Omega}} \bar{f}\,\,, \\
  R_\mathrm{pair-brem,iso}(\epsilon,\bar{\epsilon}) &= \int d\bar{\mathbf{\Omega}} R_\mathrm{pair-brem}(\epsilon,\bar{\epsilon},\mathbf{\Omega}\cdot\bar{\mathbf{\Omega}})\,\,,
  \end{aligned}
\end{equation}
where $\bar{f}_\mathrm{iso}$ depends only on energy and not on
direction. Under this assumption, the source term can be written
as
\begin{equation}
  \begin{aligned}
  \left[\frac{\partial f}{c \partial t}\right]_\mathrm{pair-brem} &= \overbrace{\int
  d\left(\frac{\bar{\epsilon}^3}{3}\right) 
    R_\mathrm{pair-brem,emis,iso}(\epsilon,\bar{\epsilon})(1-\bar{f}_\mathrm{iso})}^{\widetilde{R}_\mathrm{pair-brem,emis}}\\
    & - f \underbrace{\int d\left(\frac{\bar{\epsilon}^3}{3}\right) \left[
    R_\mathrm{pair-brem,emis,iso}(\epsilon,\bar{\epsilon})(1-\bar{f}_\mathrm{iso}) + R_\mathrm{pair-brem,abs,iso}(\epsilon,\bar{\epsilon})\bar{f}_\mathrm{iso}\right]}_{\widetilde{R}_\mathrm{pair-brem,abs}}\,\,.
  \end{aligned}
\end{equation}
\texttt{Sedonu} uses $\widetilde{R}_\mathrm{pair-brem,abs/emis}$ in the same way as $\widetilde{R}_\mathrm{abs/emis}$.

Since the NSY code evolves $f$, they use the reaction rates (units of
cm$^{-1}$) directly, but \texttt{Sedonu} needs to convert the emission reaction
rates to physical emissivities. For an emissivity $\eta$ with units of
(erg cm$^{-3}$ s$^{-1}$),
\begin{equation}
  \eta = \widetilde{R}_\mathrm{emis} \frac{\epsilon_i}{c^2h^3} \Delta\left(\frac{\epsilon^3}{3}\right)_i\,\,.
\end{equation}
Here, $\Delta\left(\epsilon^3/3\right)_i$ and $\epsilon_i$ are the
momentum-space volume (normalized by $4\pi$) and center of energy bin
$i$, respectively. The absorption and scattering reaction rates with
tildes ($\widetilde{R}$) are already equivalent to absorption and
scattering opacities.

\end{appendix}

\end{document}